\documentclass[iop]{emulateapj}

\usepackage{times,color,natbib,url,amsmath}
\usepackage{graphicx,float,psfrag}
\usepackage{tabularx}
\usepackage{latexsym,amsmath,amssymb}
\usepackage{natbib}
\usepackage{verbatim}
\usepackage{multirow}

\bibpunct{(}{)}{;}{a}{}{,}

\newcommand {\asca} {{\it ASCA}}

\newcommand {\xmm} {\textsl{XMM-Newton}}
\newcommand {\chandra} {\textsl{Chandra}}

\newcommand {\nustar} {\textsl{NuSTAR}}

\def \rsun {\ifmmode$R$_{\odot}\else R$_{\odot}$}

\def \hcm {\hbox {\ifmmode $ atoms cm$^{-2}\else atoms cm$^{-2}$\fi}}

\def\approxgt{\mathrel{\hbox{\rlap{\lower.55ex \hbox {$\sim$}}
        \kern-.3em \raise.4ex \hbox{$>$}}}}
\def\approxlt{\mathrel{\hbox{\rlap{\lower.55ex \hbox {$\sim$}}
        \kern-.3em \raise.4ex \hbox{$<$}}}}

\newcommand {\degree} {$^{\circ}$}
\def \arcmin {\hbox{$^\prime$}}
\def \arcsec {\hbox{$^{\prime\prime}$}}

\def \src {GRO~J1744$-$28}

\begin{document}

\title{Simultaneous \nustar/\chandra\ observations of the Bursting
  Pulsar GRO~J1744-28 during its third reactivation}

%
% Authors
%
\author{G.~Younes$^{1,2}$, C.~Kouveliotou$^{3,2}$,
  B.~W.~Grefenstette$^{4}$, J.~A.~Tomsick$^5$, A.~Tennant$^{3,2}$,
  M.~H.~Finger$^{1,2}$, F.~F\"urst$^4$, K.~Pottschmidt$^{6,7}$,
  V.~Bhalerao$^8$, S.~E.~Boggs$^5$, L.~Boirin$^9$,
  D.~Chakrabarty$^{10}$, F.~E.~Christensen$^{11}$, W.~W.~Craig$^{5,12}$,
  N.~Degenaar$^{13}$, A.~C.~Fabian$^{14}$, P.~Gandhi$^{15}$, E.~G\"o\u{g}\"u\c{s}$^{16}$,
  C.~J.~Hailey$^{17 }$, F.~A.~Harrison$^4$, J.~A.~Kennea$^{18}$,
  J.~M.~Miller$^{11}$, D.~Stern$^{19}$, W.~W.~Zhang$^{20}$}

 \affil{
$^1$ Universities Space Research Association, 6767 Old Madison Pike, Suite 450, Huntsville, AL 35806, USA \\
$^2$ NSSTC, 320 Sparkman Drive, Huntsville, AL 35805, USA \\
$^3$ Astrophysics Office, ZP 12, NASA-Marshall Space Flight Center, Huntsville, AL 35812, USA \\
$^4$ Cahill Center for Astrophysics, 1216 East California Boulevard, California Institute of Technology, Pasadena, California 91125, USA\\
$^5$ Space Sciences Laboratory, 7 Gauss Way, University of California, Berkeley, CA 94720-7450, USA\\
$^{6}$ Center for Space Science and Technology, University of Maryland Baltimore County, Baltimore, MD 21250, USA\\
$^{7}$ CRESST and NASA Goddard Space Flight Center, Astrophysics Science Division, Code 661, Greenbelt, MD 20771, USA\\
$^8$ Inter-University Center for Astronomy and Astrophysics, Post Bag 4, Ganeshkhind, Pune 411007, India\\
$^9$ Observatoire Astronomique de Strasbourg, 11 Rue de l'Universit\'e, 67000 Strasbourg France\\
$^{10}$ Kavli Institute for Astrophysics and Space Research, Massachusetts Institute of Technology, Cambridge, MA 02139, USA\\
$^{11}$ DTU Space, National Space Institute, Technical University of Denmark, Elektrovej 327, DK-2800 Lyngby, Denmark\\
$^{12}$ Lawrence Livermore National Laboratory, Livermore, CA 94550, USA\\
$^{13}$ Department of Astronomy, University of Michigan, 1085 South University Avenue, Ann Arbor, MI 48109, USA\\
$^{14}$ Institute of Astronomy, Madingley Road, Cambridge CB3 0HA\\
$^{15}$ Department of Physics, Durham University, South Road, Durham DH1 3LE, UK\\
$^{16}$ Sabanc\i~University, Orhanl\i-Tuzla, \.Istanbul 34956, Turkey \\
$^{17}$ Columbia Astrophysics Laboratory, Columbia University, New York, NY 10027, USA\\
$^{18}$ Department of Astronomy \& Astrophysics, The Pennsylvania State University, 525 Davey Lab, University Park, PA 16802, USA\\
$^{19}$ Jet Propulsion Laboratory, California Institute of Technology, Pasadena, CA 91109, USA\\
$^{20}$ NASA Goddard Space Flight Center, Greenbelt, MD 20771, USA
}

%\date{}

\begin{abstract}

We report on a 10~ks simultaneous \chandra/HETG$-$\nustar\ observation
of the Bursting Pulsar, \src, during its third detected outburst since
discovery and after nearly 18 years of quiescence. The source is
detected up to 60~keV with an Eddington persistent flux level. Seven
bursts, followed by dips, are seen with \chandra, three of which are
also detected with \nustar. Timing analysis reveals a slight increase
in the persistent emission pulsed fraction with energy (from $10$\% to
$15$\%) up to 10~keV, above which it remains constant. The
$0.5-70$~keV spectra of the persistent and dip emission are the same
within errors, and well described by a blackbody (BB), a power-law
with an exponential rolloff, a 10~keV feature, and a 6.7~keV emission
feature, all modified by neutral absorption. Assuming that the BB
emission originates in an accretion disc, we estimate its inner
(magnetospheric) radius to be about $4\times10^7$~cm, which translates
to a surface dipole field $B\approx9\times10^{10}$~G. The
\chandra/HETG spectrum resolves the 6.7~keV feature into
(quasi-)neutral and highly ionized Fe~XXV and Fe~XXVI emission
lines. XSTAR modeling shows these lines to also emanate from a
truncated accretion disk. The burst spectra, with a peak flux more
than an order of magnitude higher than Eddington, are well fit with a
power-law with an exponential rolloff and a 10~keV feature, with
similar fit values compared to the persistent and dip spectra. The
burst spectra lack a thermal component and any Fe features. Anisotropic
(beamed) burst emission would explain both the lack of the BB and any
Fe components.

\end{abstract}


\section{Introduction}
\label{Intro}

\src\ is a high-energy transient in a Low-Mass X-ray Binary (LMXB)
system, and only the second source, besides the Rapid Burster
\citep{lewin93ssr}, observed to emit multiple type II
X-ray bursts,  i.e., due to spasmodic accretion rather
than thermonuclear burning. The source was discovered in 1996 with the Burst
and Transient Source Experiment (BATSE) on board the {\sl Compton
  Gamma Ray Observatory   (CGRO)}, when it emitted a series of hard
X-ray bursts during a period lasting $\sim150$ days
\citep{kouveliotou96:1744}. Soon after its discovery,
\citet{finger96Natur:1744} reported that the timing properties of the
persistent X-ray emission pointed towards a magnetized neutron star
pulsating at 2.14~Hz, accreting material from a low-mass companion in
a nearly circular orbit with an orbital period of 11.8~days.  At that
time \src\ was the first source to show bursts and pulsations, hence
the source was nicknamed ``the Bursting Pulsar'' (hereafter
BP). The BP emerged from quiescence again almost exactly one year
after its first outburst, in December 1996
\citep{woods99ApJ:1744}. This second  outburst was very similar to the
first including both burst and persistent X-ray emission
characteristics \citep{woods99ApJ:1744}.

The BP and its two outbursts were studied extensively during the first
few years after its discovery.  The X-ray bursts from the source were
classified as type II bursts, based on their spectra, energetics
\citep{kouveliotou96:1744}, and their resemblance to the bursts
observed from the Rapid Burster \citep{lewin96ApJ:1744}. Type II
bursts are most likely the result of  some sort of instability (whose
origin is still unknown) in the accretion disk resulting in the onset
of mass inflow onto the neutron star which is responsible for the
bursting activity. The BP average burst duration was 10~s and each
burst was followed by a dip in flux below that of the pre-burst
persistent emission. The flux recovered exponentially back to the
pre-burst persistent emission level on time-scales of a few hundred
seconds \citep{giles96ApJ:1744,strickman96ApJ:1744,aptekar98:1744II,aptekar98:1744I,borkus97:1744,aleksandrovich98:1744,woods99ApJ:1744,
  mejia02ApJ:1744}. Pulsations at the spin frequency of the source
were also detected during bursts, albeit with an average time lag of
about 50 ms compared to the pre-burst pulses \citep{
  strickman96ApJ:1744,stark96ApJ:1744,koshut98ApJ:1744,
  woods00ApJ:1744}. The pre-burst pulse profile was subsequently
recovered on timescales of a few hundred seconds \citep{
  stark96ApJ:1744}. \citet{miller96ApJ:1744} attributed these lags to
the accretion column geometry at the pole.

There is no direct estimate as yet of the magnetic field of the
BP. \citet[][see also \citealt{daumerie96Natur:1744}]{
  finger96Natur:1744} placed an upper limit on the dipole magnetic
field of $B\lesssim6\times10^{11}$~G based on the spin-up rate of the
source and the persistent pulsed luminosity. \citet{
  rappaport97ApJ:1744} deduced from binary evolution calculations that
the dipole magnetic field of \src\ lies in the range  of
$(1.8-7.0)\times10^{11}$~G, with a most probable value of
$2.7\times10^{11}$~G. Finally, \citet{cui1997ApJ:1744gx1p4} derived a
surface magnetic field of $B\approx2.4\times10^{11}$~G, assuming that
the propeller effect is the reason for  the non-detection of X-ray
pulsations when the source persistent flux dropped below a certain
level. It is, therefore, likely that the surface magnetic field of the
BP lies between classical X-ray accreting pulsars ($\sim10^{12}$~G)
and LMXBs ($\sim10^{9}$~G). This intermediate strength surface field
could be an important parameter defining the unusual properties of
this source; hence, the determination of its exact value is of crucial
importance.

\citet{nishiuchi99ApJ:1744} studied the $0.5-10$\,keV spectrum of
\src\ during outbursts using \asca. They found a spectrum well
described by an absorbed power-law (PL) and line-like emission between 6
and 7~keV, most likely from Fe reprocessed in the accretion disk. The
heavy absorption towards the source ($N_{\rm
  H}\approx10^{23}$~cm$^{-2}$) places the BP at the Galactic
center, likely at 8~kpc. 

The quiescent X-ray counterpart of the BP was discovered with
\chandra\ \citep{Wijnands02apj:1744} and was confirmed one month later
with \xmm\ \citep{daigne02AA:1744}. The spectrum in quiescence is
soft and could be fit with either a PL model with $\Gamma=2-5$ or a
blackbody with $kT=0.4-1$~keV, implying a quiescent $0.5-10$~keV X-ray
luminosity of $3\times10^{33}$~erg~s$^{-1}$ at 8~kpc. Using the
\chandra\ position, \citet[][see also \citealt{cole97ApJ:1744,
  augustinjn97ApJ:1744}]{gosling07MNRAS:1744} found two potential
infrared counterparts within the BP error circle, with the most likely
candidate being a giant star of type G4~III.

On 2014 January 18, the {\it Monitor of All-sky X-ray Image (MAXI})
Gas Slit Camera detected enhanced hard X-ray emission from the
Galactic center region \citep{negoro14ATel:1744}. Following the
detection, they examined archival data from the {\it Swift} Burst
Alert Telescope (BAT), and found that the X-ray emission from the BP
had increased compared to its quiescent level. Soon after, the source
triggered BAT on 2014 January 18 \citep{negoro14ATel:1744}.
\citet{finger14ATel:1744} detected pulsations from the direction of
\src\ at the 2.14~Hz spin period of the source during January
$19.0-21.0$ using  the {\it Fermi} Gamma-ray Burst Monitor. Finally,
the {\sl Swift} X-ray Telescope (XRT) observed the BP on 2014 February
2 \citep{kennea14ATel:1744}, detecting a bright source at the
\chandra\ location, confirming that the source entered a new outburst
after about 18~years of quiescence \citep[see also
][]{antonino14ATel:1744,linares14ATel:1744,chakrabarty14ATel:1744,pintore14ATel:1744,pandey14ATel:1744,sanna14ATel:1744,negoro14ATel63:1744}. \citet{masetti14ATel99:1744}
discovered infrared brightening of the G4 III candidate counterpart 
contemporary with the X-ray outburst, confirming its identification
as the BP companion.

Here we report our results of the analysis of a 10~ks simultaneous
\chandra\ and \nustar\ observation of the BP taken on 2014 March
3. Section~\ref{obs} describes  the observations and data reduction
techniques. Our results are presented in Section~\ref{res}, and
discussed in Section~\ref{discuss}.

\section{Observations and data reduction}
\label{obs}

\subsection{Chandra}
\label{chanobs}

We observed the BP with \chandra\ using the High Energy Transmission
Grating (HETG) in continuous clocking mode (CC-mode) with all six CCDs
of the ACIS-S array. The HETG comprises two sets of gratings, the
medium energy grating (MEG), operating in the energy range of
$0.4-7$~keV, and the high energy grating (HEG) with energy coverage in
the range of $0.8-10$~keV, and a spectral resolution (FWHM)
$\Delta E=0.4-77$~eV. Each grating spectrum is dispersed
along the ACIS-S CCDs into positive and negative spectral orders. In
addition, each grating observation results in an on-axis undispersed
image with the CCD spectral resolution. We used the CC-mode to obtain
the highest possible temporal resolution of 2.85~ms, at the expense of
obtaining a one dimensional image of the source.

The observation took place on 2014 March 3, 08:59:06 UTC, with 10~ks
of good time intervals (GTI). A comparison between the zeroth order
and the dispersed HEG~$\pm$ first order light curve reveals that the
bursts are completely missing from the zeroth order light curve due to
heavy pile-up. The source persistent emission also suffered a
$10\%$ pile-up effect in the zeroth order. On the other 
hand, the dispersed grating spectra have a much lower total count
rate compared to that of the zeroth order spectra. This results in
spectra free of pile-up except during the peak of the bursts when
pile-up still occurred at the $10\%$ level.

To use the grating arms in our timing analysis, the photons assigned
times needed to be corrected for their diffraction angle, which is
directly proportional to the grating time offset with respect to the
zeroth order.  This time offset, $\delta t$, relative to the
zeroth order location is

\begin{equation}
\delta t=-\frac{\sin(tg\_r_{\rm i})\times X_{\rm R}\times\sin\alpha_{\rm i}}{\Delta_{\rm p}}~t_{\rm p}~[{\rm s}]
\end{equation}

\noindent where $tg\_r_{\rm i}$ is the diffraction angle of each
photon $i$, $X_{\rm R}$ is the Rowland spacing, $\alpha_{\rm i}$ is
the grating clocking angle, $\Delta_{\rm p}$ is the pixel size, and
$t_{\rm p}$ is the read time per row (2.85~ms). 

The use of CC-mode with any \chandra\ grating observation introduces
some complications to the data reduction and
analysis\footnote{http://cxc.harvard.edu/cal/Acis/Cal\_prods/ccmode/ccmode\_final\_doc02.pdf}. For
instance, due to the fact that there is no spatial information (the
clocking rows are all collapsed into one pixel), the soft X-ray
background, usually low in TE mode, is enhanced by 3 orders of
magnitude.

\begin{figure}[t]
\begin{center}
\includegraphics[angle=0,width=0.45\textwidth]{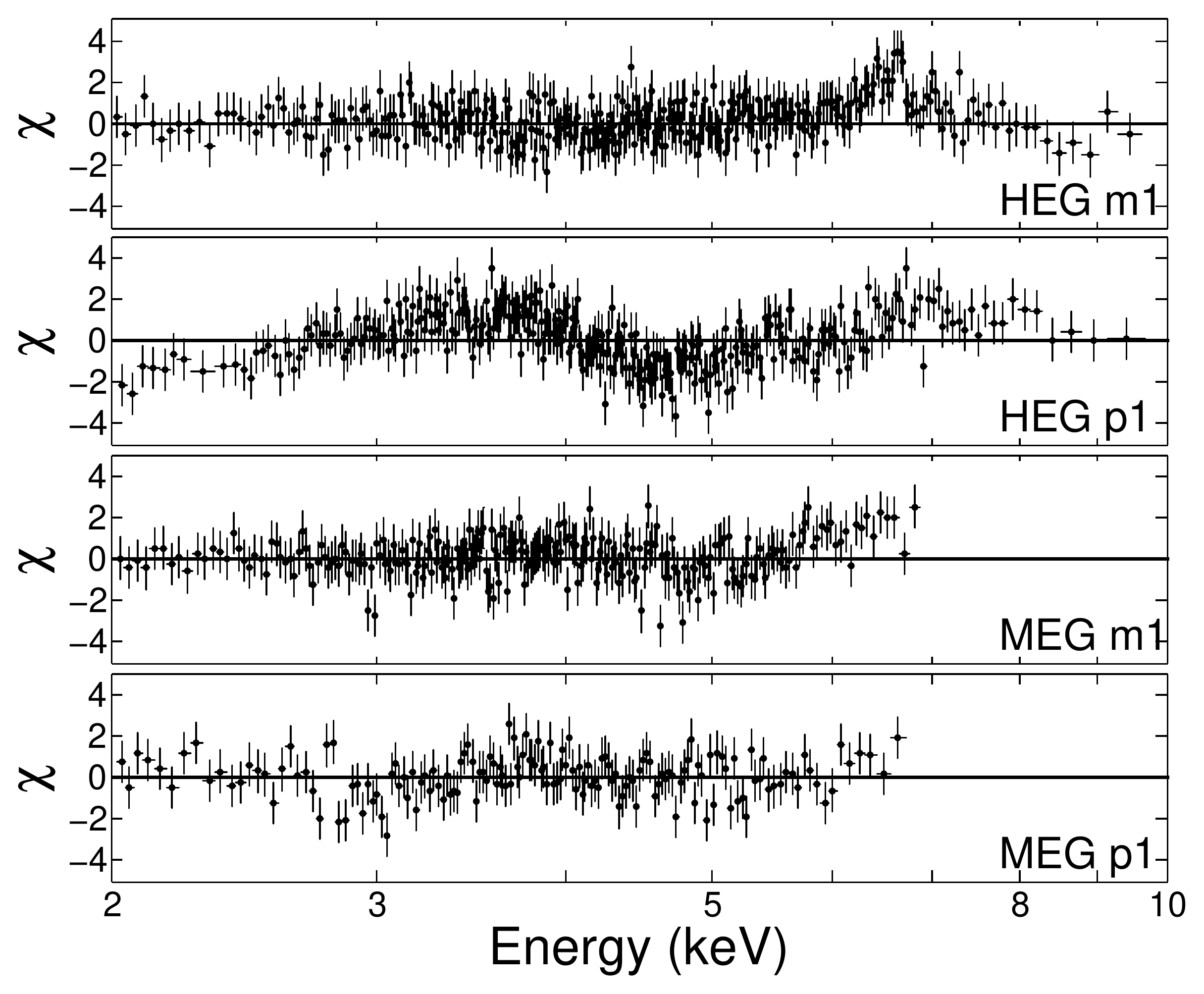}\\
\caption{Residuals of an absorbed PL fit to the non-burst
  emission interval for the $+$ and $-$ arms of both the HEG and MEG 
  1st order spectra. The wiggle seen in the HEG~p1 spectrum between
  2 and 6~keV is not detected in any of the other three spectra. We
  conclude that this shape is artificial and refrain from using the
  HEG~p1 in the non-burst spectral analyses.}
\label{calibration}
\end{center}
\end{figure}

A more pressing issue, that can alter grating dispersed spectra in
CC-mode, is the dust scattering halo, usually present around bright
absorbed sources. The BP resides in the Galactic center region, hence,
it is heavily absorbed, with a hydrogen column density $N_{\rm
  H}\approx10^{23}$~cm$^{-2}$. The brightness of the source
produces a diffuse scattering halo, the emission from which disperses
and blends with the  source dispersed spectrum. The significance of this
effect depends on the incident source spectrum, with hard sources
affected less than soft ones. Luckily, the BP has a hard X-ray spectrum
(Section~\ref{specana}) which reduces the impact of this background on
the source spectrum.

Since no spatial information exists when using CC-mode, we extract
the MEG and HEG backgrounds using the order sorting plots, which
display the energies of the dispersed events versus the ratios of
these energies over the event positions on the grating arm$^1$.
On-axis point-source photons should distribute tightly and
symmetrically around the extraction order, while diffuse photons have
a larger scatter. Finally, the extracted background is normalized to
the excluded source region.  These backgrounds are used for both
timing and spectral analyses.

\begin{figure}[t!]
\begin{center}
\includegraphics[angle=0,width=0.45\textwidth]{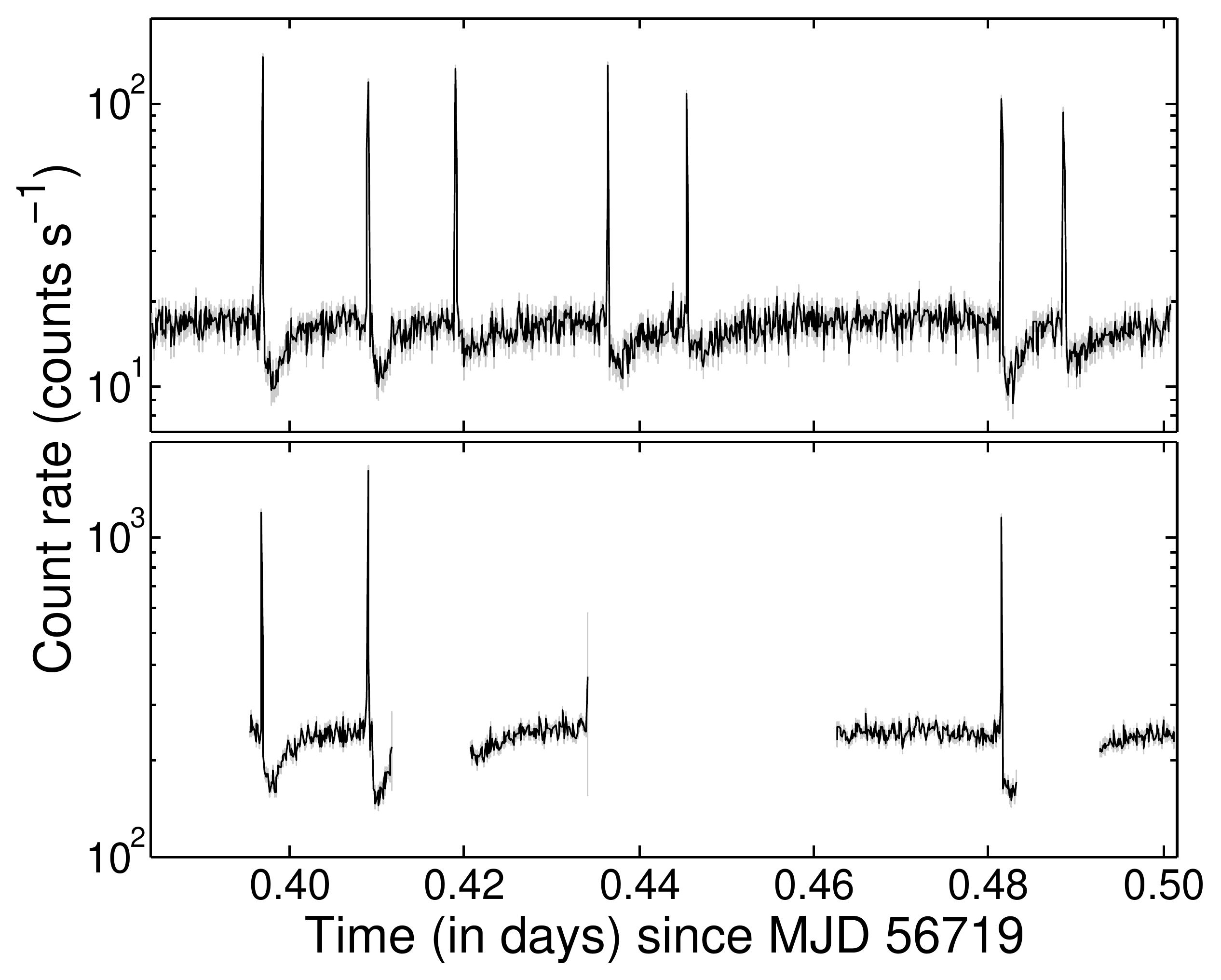}\\
\caption{\chandra\ ({\sl top panel}) and {\it NuSTAR} ({\sl bottom panel})
  light curves in the 2-10 and 3-10 keV energy ranges,
  respectively. Both light curves are binned with 10~s time resolution. A total
  of three bursts are detected simultaneously by \chandra\ and
  \nustar.}
\label{chanNusLC}
\end{center}
\end{figure}

\begin{figure}[t!]
\begin{center}
\includegraphics[angle=0,width=0.45\textwidth]{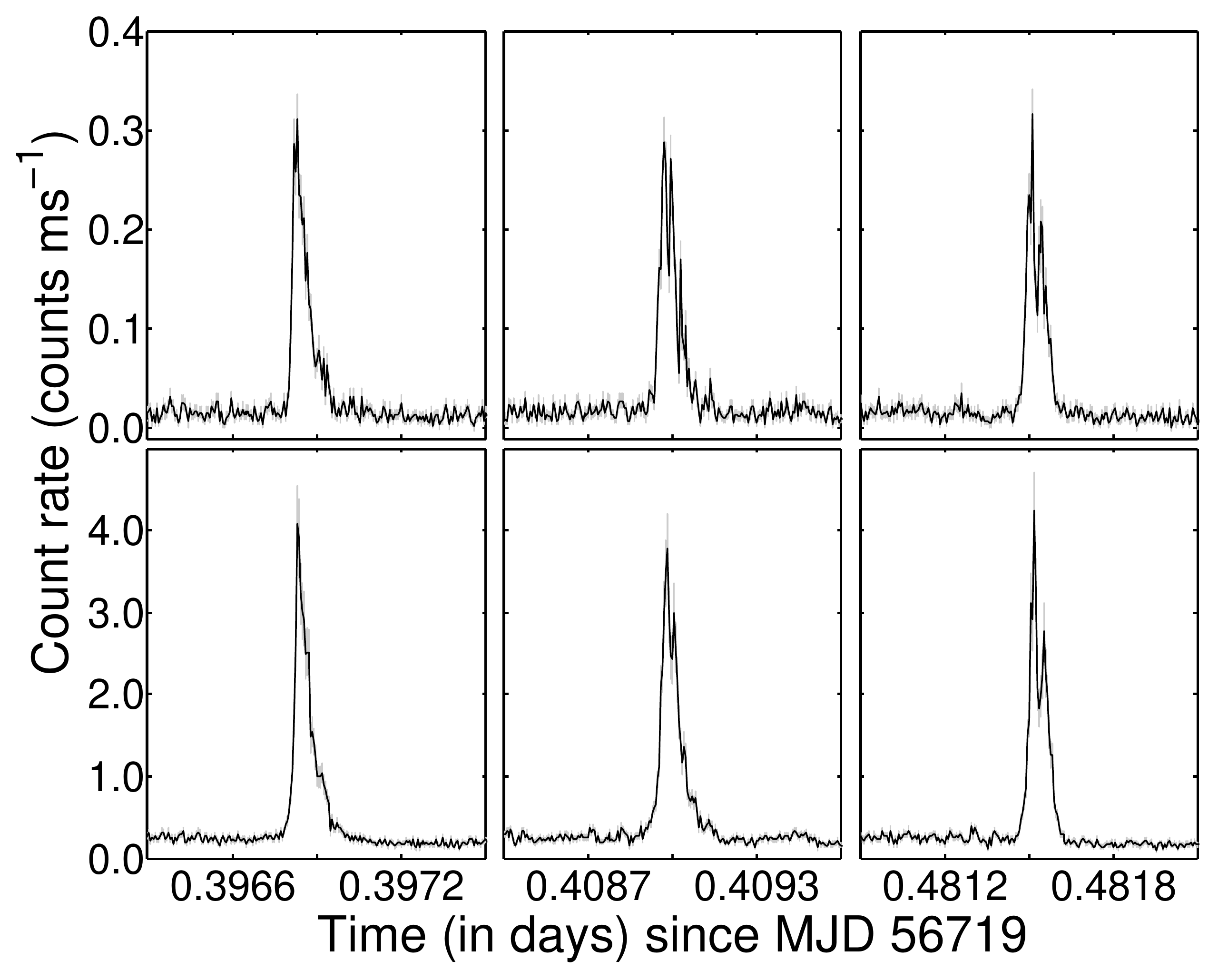}
\caption{Light curves of the three bursts that are covered
  simultaneously by \chandra\ ({\sl top panel}) and {\it NuSTAR}\
  ({\sl bottom panel}) in the $2-10$ and $3-10$~keV energy 
  ranges, respectively, plotted with 0.5~s time resolution.}
\label{chanNusLCBursts}
\end{center}
\end{figure}

In addition to the background complications when dealing with CC-mode
observations, there are calibration uncertainties$^1$ between the
different orders, e.g., complicated charge transfer inefficiency (CTI)
corrections on the events. To check for potential differences, all
\chandra\ analyses were initially performed on the separate HEG and MEG
arms. Temporal analysis returned consistent results between all the
different arms. Spectral analyses, on the other hand, showed that the
HEG~p1 (hereafter, p1 refers to the positive and m1 to the negative
first order grating arms) spectrum is markedly different from the
rest, i.e., MEG~p1, MEG~m1, and HEG~m1. Figure~\ref{calibration} shows
a PL fit to the different spectral arms, where a wiggle between 2 and
6~keV is present only in the HEG~p1 spectrum. We conclude that this
feature is not real, and most likely due to either miscalibration, an
improper modeling of the background, or both. Hence, the HEG~p1 arm
is excluded from the spectral analyses, except for bursts, where the
above feature is not present (likely due to the small integration
times during bursts and/or the fact that the emission during burst
intervals includes minimal background).

All narrow features in the \chandra\ spectra are seen above 6~keV. Due
to the lower spectral resolution and collecting area of the MEGs at
energies $\gtrsim5$~keV, we also exclude these spectra in the
analysis. For our timing analyses, we use the HEG first order gratings
(positive and negative arms combined, i.e., HEG~1).

\subsection{NuSTAR}

The {\sl Nuclear Spectroscopic Telescope Array} (\nustar) is a NASA
Small Explorer (SMEX) satellite launched on 2012 June 13
\citep{harrison13ApJ:NuSTAR}. It is the first orbiting focusing hard
X-ray telescope, observing the sky in an energy range from 3 to 79 keV
with two co-aligned X-ray optics which focus X-rays onto two
independent detector planes (FPMA and FPMB), each composed of four
CdZnTe detectors. The field of view of \nustar\ is roughly 12\arcmin\
x 12\arcmin\, with a point-spread function with a FWHM of 18\arcsec
and a half-power diameter of 58\arcsec.

\nustar\ obtained simultaneous observations of the BP during the
\chandra\ observation. The broader \nustar\ BP data set is reserved
for future work; here we concentrate on the data obtained
simultaneously with \chandra. We reduced the \nustar\ data using
NuSTARDAS v 1.3.1 and the \nustar\ CALDB 20131210, with the standard
pipeline filtering. We extracted the source photons from a circular
region with a radius of 60\arcsec; these regions were centroided
separately for FPMA and FPMB to account for the small misalignments in
the absolute aspect reconstruction for the two telescopes.

\nustar\ produces event files (e.g., each row in the event file
represents a single time-tagged photon), which we can then filter
based on the source region described above to produce ``source" event
files. 

We produced response files (ARFs and RMFs) for the \nustar\ spectral
analysis using the custom time-intervals defined in
Section~\ref{timeana}. These response files capture the response of the
instrument over the specified time ranges.

\section{Results}
\label{res}

\subsection{Temporal properties}
\label{timeana}

We show in Figure~\ref{chanNusLC} the $2-10$ keV \chandra\ HEG 1st
order light curve (top panel) of the entire 10~ks observation, with
10~s time bins. Seven bursts are detected from the source during this
observation. Following each burst, a dip in the count rate is observed,
which recovers exponentially back to the persistent level. The FPMA
light curve of the simultaneous {\it NuSTAR} observation is shown in 
the second panel in Figure~\ref{chanNusLC}, in the energy range of
$3-10$~keV and also at 10~s resolution. Only three bursts are detected
simultaneously by both \chandra\ and \nustar.
Figure~\ref{chanNusLCBursts} is a zoom in on these three bursts
plotted with a 0.5~s resolution.

\begin{figure}[t!]
\begin{center}
\includegraphics[angle=0,width=0.43\textwidth]{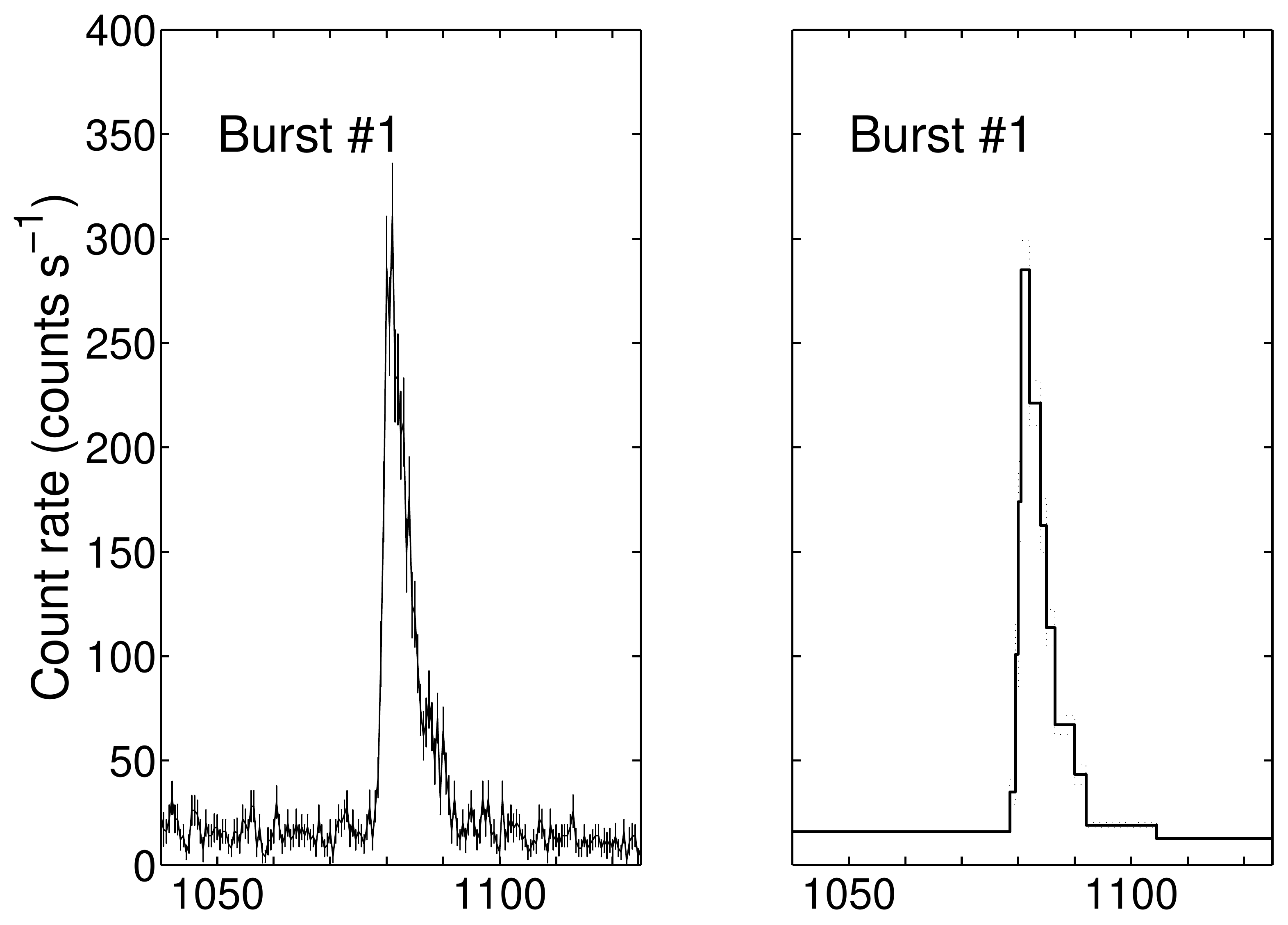}\\
\includegraphics[angle=0,width=0.43\textwidth]{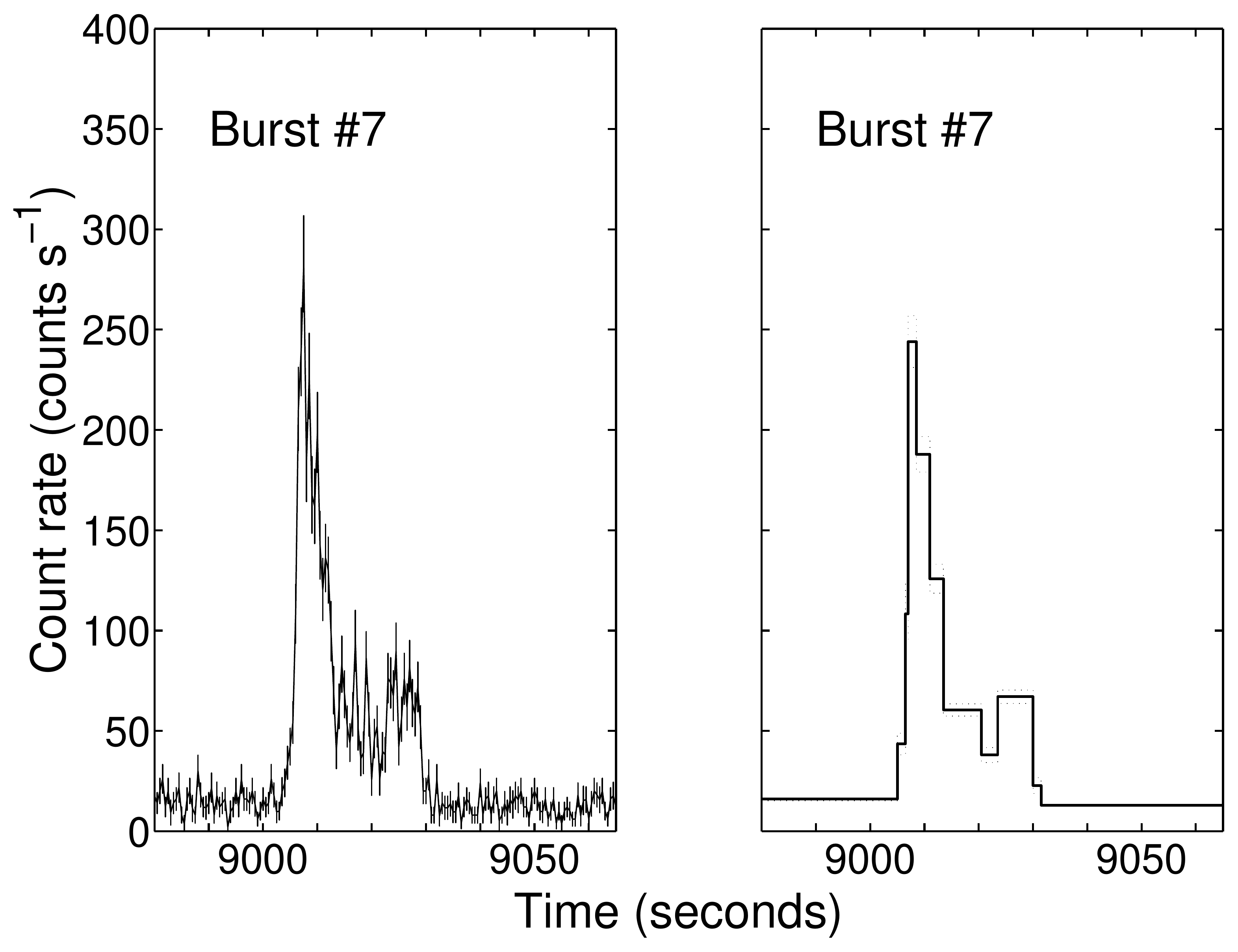}
\caption{Two examples of \chandra\ $2-10$~keV light curves of  BP bursts with
  0.5~s resolution (left panels), along with their Bayesian Blocks
  representation (right panels). Times are from the start of the \chandra\
  observation.}
\label{burstsBayes}
\end{center}
\end{figure}

We use a Bayesian Blocks algorithm \citep{scargle2013apj:BB} to
identify the beginning and end times of the 7 bursts detected with
\chandra, and to search for weaker bursts in the \chandra\ light curve
binned with 0.5~s resolution. We chose this temporal resolution as a
trade-off between speed and accuracy, considering that the source has
a comparable pulse period. Bayesian Blocks have been frequently used
for the temporal analysis of gamma-ray bursts, magnetars, and even
flares from Sgr~A$^*$
\citep[e.g.,][]{norris11ApJ:shortGRB,lin13ApJ:magnetarsBB,nowak12ApJ:sgrA,
barriere2014ApJ:sgrA}. We find that only the 7 bursts clearly visible
in Figure~\ref{chanNusLC} show significant deviations
($\gtrsim5\sigma$) from the persistent level of the source. We do not
find any weaker (mini$-$) bursts similar to the ones seen during the
first two outbursts, e.g., \citep{nishiuchi99ApJ:1744}.

The start and end times of the 7 \chandra\ bursts are recorded from
the Bayesian Blocks analysis
(Figure~\ref{burstsBayes}). Table~\ref{burstProp} shows these start
times in MJD, and durations in seconds. The uncertainty on these
durations is dominated by the temporal resolution we used, and it is
$\leq1$~s ($\leq0.5$~s error at the beginning and end of each
burst). All durations, except the last, are narrowly distributed with
a mean and 1$\sigma$ standard deviation of $12\pm2$~s. The last burst
consists of two pulses with a total duration of $25$~s
(Figure~\ref{burstsBayes}). The rise and decay times of all bursts
have a mean and 1$\sigma$ standard deviation of $4.0\pm1.0$~s, and 
$8.0\pm2.0$~s, respectively (excluding the decay time of the last
burst).

\begin{figure}[t!]
\begin{center}
\includegraphics[angle=0,width=0.45\textwidth]{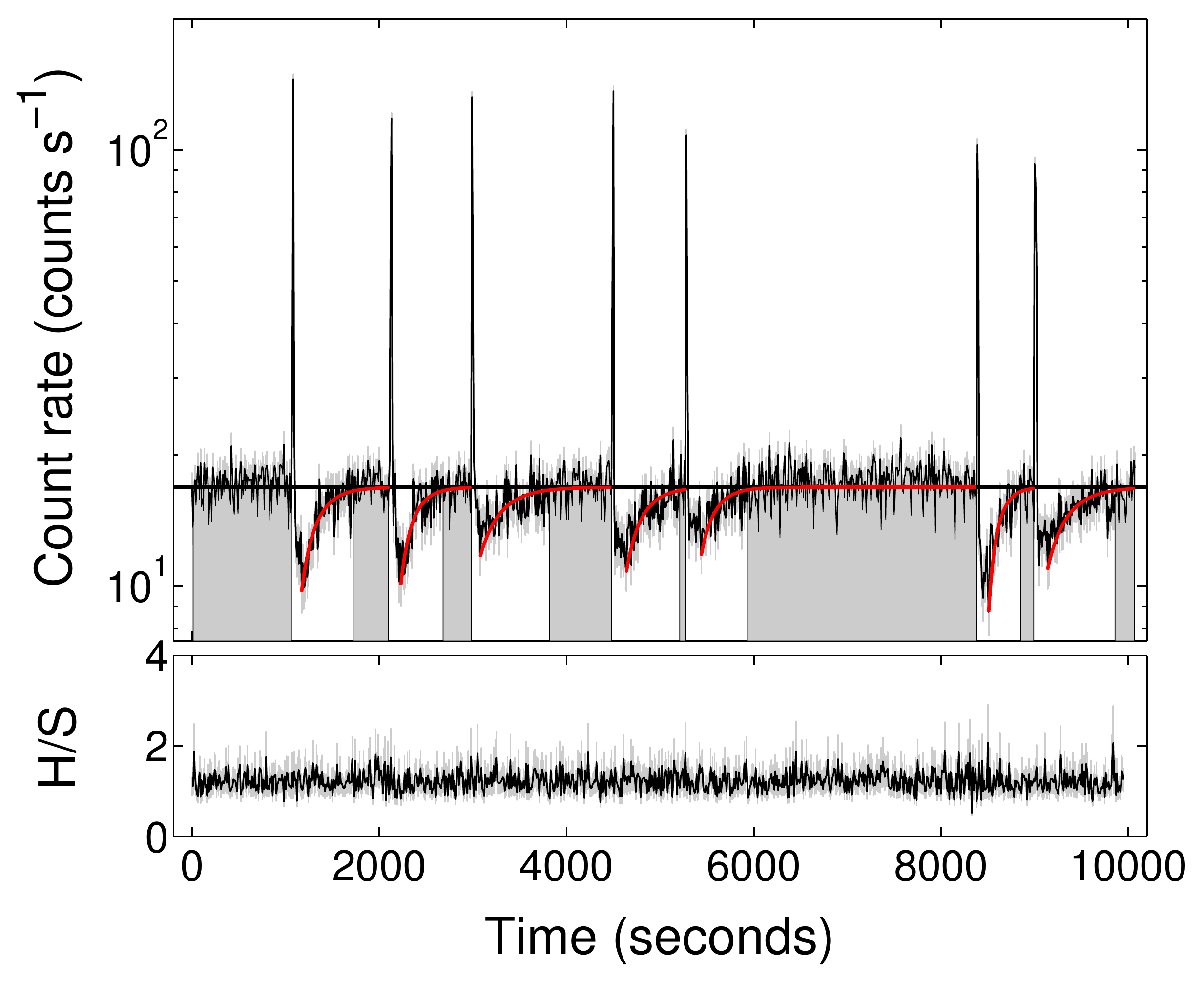}\\
\caption{{\sl Top panel.} The \chandra\ $2-10$~keV light curve with
  10~s resolution. The black solid line is the mean rate of a
  pre-defined persistent emission interval far away from bursts and
  dips. The grey areas are the persistent emission
  intervals used in our spectral analysis. The red solid curves are exponential fits to the dip
  intervals. {\sl Bottom panel.} Hardness ratio, $H/S$, evolution during the entire \chandra\ observation ($H$ is computed in the $4-6$ and $8-10$ keV range, and $S$ in the $2-4$~keV range).}
\label{fitDips}
\end{center}
\end{figure}

We identify the dip durations using the following method on the
2$-$10~keV \chandra\ light curve binned at 10~s. We search for the
time-bin with the minimum count rate immediately after the end of a
burst and up to the start of the following burst (the end of the
observation after the last burst). We then fit the light curve of each
of these time intervals with an exponential function of the form,

%-----------
% Table  1
%-----------
\begin{table*}[!t]
\caption{BP burst temporal and spectral properties.}
\label{burstProp}
\newcommand\T{\rule{0pt}{2.6ex}}
\newcommand\B{\rule[-1.2ex]{0pt}{0pt}}
\begin{center}{
\resizebox{0.95\textwidth}{!}{
\begin{tabular}{l c c c c c c c c c}
\hline
\hline
Burst \# & Start time & Duration & Rise & Decay & PL norm.$^{a}$ & Flux$^{a,d}$ & Fluence$^{a,d}$ & Luminosity$^{a,b,d}$ & Flux$_{\rm 0.5-70~keV}^{c}$\\
 \T \B & MJD & s & s & s & photons~keV$^{-1}$ cm$^{-2}$~s$^{–1}$ & $10^{-8}$~erg~cm$^{-2}$~s$^{-1}$ & $10^{-7}$~erg~cm$^{-2}$ & $10^{38}$~erg~s$^{-1}$ & $10^{-8}$~erg~cm$^{-2}$~s$^{-1}$\\
\hline
1\T &  56719.396802 &  13.5 &  2.8 &  10.8 & 8~(5$-$11)   & 8.6~(7.7$-$9.7)     & 11.6 (10.4$-$13.1) & 6.6  (5.9$-$7.4) &  10.5 (10.2$-$10.9)\\
2     &  56719.408920 &  15.5 &  4.8 &  10.8 & 7~(5$-$10)   & 7.4~(6.7$-$8.4)     & 11.5 (10.4$-$13.0) & 5.7  (5.1$-$6.4) &  9.3 (8.9$-$9.6)\\
3     &  56719.418949 &  12.5 &  5.0   &  7.5     & 8~(6$-$12)   & 9.2~(8.3$-$10.4)   & 11.5 (10.4$-$13.0) & 7.1  (6.4$-$8.0) & $---$ \\
4     &  56719.436373 &  9.0   &  3.5   &  5.5     & 10~(7$-$14) & 10.7~(9.7$-$12.1) & 9.7 (8.7$-$10.9) & 8.2 (7.4$-$9.3) & $---$ \\
5     &  56719.445465 &  11.0 &  3.5   &  7.5     & 7~(5$-$10)   & 7.3~(6.5$-$8.3)     & 8.3 (7.2$-$9.1) & 5.6  (5.0$-$6.4)  & $---$ \\
6     &  56719.481460 &  12.5 &  5.3 &  7.3   & 7~(5$-$11)   & 8.2~(7.4$-$9.2)     & 10.2 (9.2$-$11.5) & 6.3  (5.6$-$7.1) &  9.4 (9.0$-$9.7)\\
7     &  56719.488544 &  25.0 &  2.3 &  22.3 & 5~(3$-$7)     & 5.3~(4.8$-$5.9)     & 13.1 (11.9$-$14.8) & 4.0  (3.7$-$4.5) & $---$ \\
\hline
\end{tabular}}}
\end{center}
\begin{list}{}{}
\item[{\bf Notes.}] $^a$ Absorbed PL with $N_{\rm
    H}=9.0\times10^{22}$~cm$^{-2}$, and $\Gamma=1.2$. $^b$ Assuming a
  distance of 8~kpc. $^c$ Combined \chandra\ and \nustar\ data fit
  with an absorbed {\sl cutoffPL} with $N_{\rm
    H}=(10.0\pm1.0)\times10^{22}$~cm$^{-2}$, $\Gamma=0.5\pm0.1$, and
  $E_{\rm fold}=8.8\pm0.7$~keV. $^d$ Derived in the energy range
  $0.5-10$~keV.
\end{list}
\end{table*}
%-----------
% Table  1
%-----------

%-----------
% Table  2
%-----------
\begin{table*}[!t]
\caption{BP dip temporal and spectral properties.}
\label{dipProp}
\newcommand\T{\rule{0pt}{2.6ex}}
\newcommand\B{\rule[-1.2ex]{0pt}{0pt}}
\begin{center}{
\resizebox{0.95\textwidth}{!}{
\begin{tabular}{l c c c c c c c c}
\hline
\hline
Dips \# & $t_{\rm min}^{a}$ & $\tau^{b}$ & $T_{\rm S\_dip}^{c}$ & $F_{\rm avr, t\_min}^d$ & $F_{\rm t\_min}^{e}$ & Flux$_{\rm total}$ & Fluence$^f$ & Flux$_{\rm 0.5-70~keV}^{g}$\\
 \T \B & MJD & s & s & $10^{-9}$~erg~cm$^{-2}$~s$^{-1}$ & $10^{-9}$~erg~cm$^{-2}$~s$^{-1}$ & $10^{-8}$~erg~cm$^{-2}$~s$^{-1}$ & $10^{-6}$~erg~cm$^{-2}$ & $10^{-8}$~erg~cm$^{-2}$~s$^{-1}$ \\
\hline
1\T &  56719.397861 & $182\pm14$ & 78   & 7.9~(6.9$-$9.2)   & $6.8\pm0.3$ & 1.00 (0.98-1.03) & 1.3 (1.5-1.1) & 2.2 (2.1$-$2.4)\\
2     &  56719.410129 & $147\pm15$ & 89   & 8.0~(7.0$-$9.3)   & $7.1\pm0.2$ & 1.03 (1.00-1.05) & 1.0 (0.9-1.2) & $---$ \\
3     &  56719.419967 & $246\pm27$ & 75   & 9.7~(8.6$-$11.3) & $8.2\pm0.3$ & 1.09 (1.07-1.11) & 1.1 (0.9-1.3) & $---$ \\
4     &  56719.438023 & $189\pm23$ & 133 & 8.6~(7.6$-$10.0) & $7.3\pm0.3$ & 1.02 (1.00-1.04) & 1.4 (1.2-1.6) & $---$ \\
5     &  56719.447282 & $162\pm25$ & 145 & 9.1~(8.1$-$10.6) & $8.0\pm0.3$ & 1.06 (1.04-1.08) & 1.0 (0.8-1.2) & $---$ \\
6     &  56719.482814 & $112\pm12$ & 104 & 7.8~(6.9$-$9.1)   & $6.1\pm0.3$ & 0.95 (0.94-0.98) & 1.2 (1.3-1.1) & $---$ \\
7     &  56719.490106 & $239\pm22$ & 109 & 9.0~(7.9$-$10.4) & $7.4\pm0.3$ & 1.03 (1.01-1.05) & 1.5 (1.3-1.7) & $---$ \\
\hline
\end{tabular}}}
\end{center}
\begin{list}{}{}
\item[{\bf Notes.}] $^a$ Dip times at minimum count rate after
  burst. $^b$ Dip characteristic time-scale for recovery. $^c$
  Duration of interval from end of burst to $t_{\rm min}$. $^d$
  Calculated in an 80~s  interval centered on $t_{\rm min}$. $^e$
  Calculated at $t_{\rm min}$ over 10~s by converting count rates to
  fluxes with PIMMS. Errors reflect the count rate errors only. $^f$
  Fluence deficiency in the dip. $^{g}$ Combined \chandra\ and
  \nustar\ data.
\end{list}
\end{table*}
%-----------
% Table  2
%-----------

\begin{equation}
\label{equation1}
C(t) = (C_{\rm p}-C_{\rm min})(1-\exp\{-(t-t_{\rm min})/\tau\})+C_{\rm min};
\end{equation}

\noindent where $C_{\rm min}$ is the minimum count rate at time $t_{\rm min}$,
$C_{\rm p}$ is the persistent count rate level, and $\tau$ is the
characteristic time-scale representing 63.2\% recovery of the count
rate to $C_{\rm p}$. We first performed fits keeping both $C_{\rm p}$
and $\tau$ as free parameters. We find that $C_{\rm p}$ is similar
in all cases with good enough coverage after the recovery, hence we
keep $C_{\rm p}$ constant at the mean count rate value calculated from
time intervals far away from bursts and dips (black solid line in
Figure~\ref{fitDips}). Figure~\ref{fitDips} shows in red our
exponential fits to the dips, and Table~\ref{dipProp} lists the dip
temporal properties. We find that $\tau$ ranges between $112$ and
$246$~s, with a mean and 1$\sigma$ standard deviation of $191\pm43$~s,
whereas the average time from the end of a burst to the minimum count
rate of the dip, $T_{\rm S\_dip}$, is $105\pm27$~s. We note that no
bursts are seen during the dipping intervals, i.e., all bursts are
emitted after the dip recovered to at least the $95\%$ level of the
persistent emission. The grey areas in Figure~\ref{fitDips} represent
the persistent emission time intervals excluding bursts and dip
intervals. 

Finally, we searched for any strong spectral variations in the
\chandra\ observation, especially during bursts, by looking at the
evolution of the source flux hardness ratio, $H/S$, where $H$ includes the
energy ranges $4-6$ and $8-10$~keV (to avoid contamination from the Fe
line complex, see Section~\ref{specana}), and $S$ includes the $2-4$~keV
range. The bottom panel of Figure~\ref{fitDips} shows the $H/S$ derived
from light curves with a 0.5~s resolution during bursts and with 10~s
bins elsewhere. We do not find any spectral variations during bursts
in the \chandra\ observation compared to the non-burst emission, at
the above temporal resolution.

\subsection{Pulse Profile analysis}

\begin{figure*}[th!]
\begin{center}
\includegraphics[angle=0,width=0.9\textwidth]{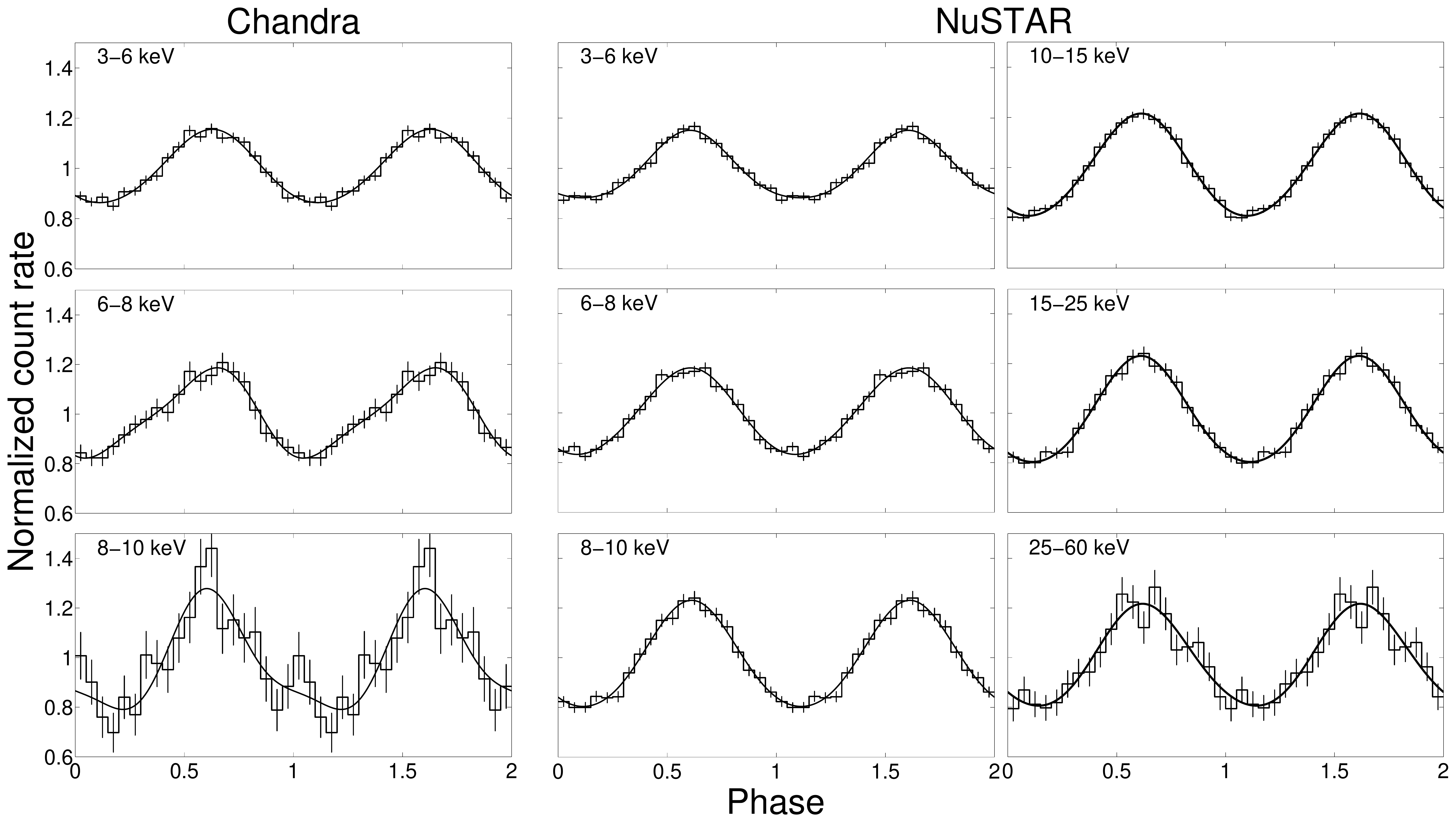}
\caption{The persistent emission pulse profiles in different energy
  bands as measured with \chandra\ ({\sl left column}) and
  \nustar ({\sl middle and right columns}). The black solid lines are the best fit
  models at the pulse period of \src.}
\label{PP_persisNU}
\end{center}
\end{figure*}

For each of the intervals defined in the previous section, we first
apply a barycenter correction for the \chandra\ and \nustar\
time-tagged events (see also Section~\ref{chanobs} for a description
of the \chandra\ grating time correction). We then correct these
times for the binary motion of the system, using the orbital
parameters provided by the GBM pulsar
team\footnote{http://gammaray.nsstc.nasa.gov/gbm/science/pulsars/}. We
estimate the spin frequency of the pulsar from the persistent data 
by locating the peak Rayleigh power, $nR^2$, in a frequency range
expected to contain the spin frequency. The Rayleigh power is given by

\begin{equation} 
nR^2 = {1 \over n} | \sum_{i = 0}^{n-1} \exp\{2\pi\nu t_i\} |^2
\end{equation}

\noindent \citep{brazier94MNRAS:pulsations}, where $n$ is the number
of events, $\nu$ a trial frequency, and $t_i$ a barycenter and binary
corrected event time. The peak power of 891 occurs at frequency
2.1411203(16)~Hz. The one sigma error is determined by the change in
frequency required for the Rayleigh power to drop by 0.5. Finally, we
epoch-fold the data at the spin frequency derived above to compute a
pulse profile (PP). For the persistent and dip intervals, we extract
\chandra\ and \nustar\ pulse profiles in different energy bands,
chosen to have comparable number of events
(Figure~\ref{PP_persisNU}).

\begin{figure}[th!]
\begin{center}
\includegraphics[angle=0,width=0.45\textwidth]{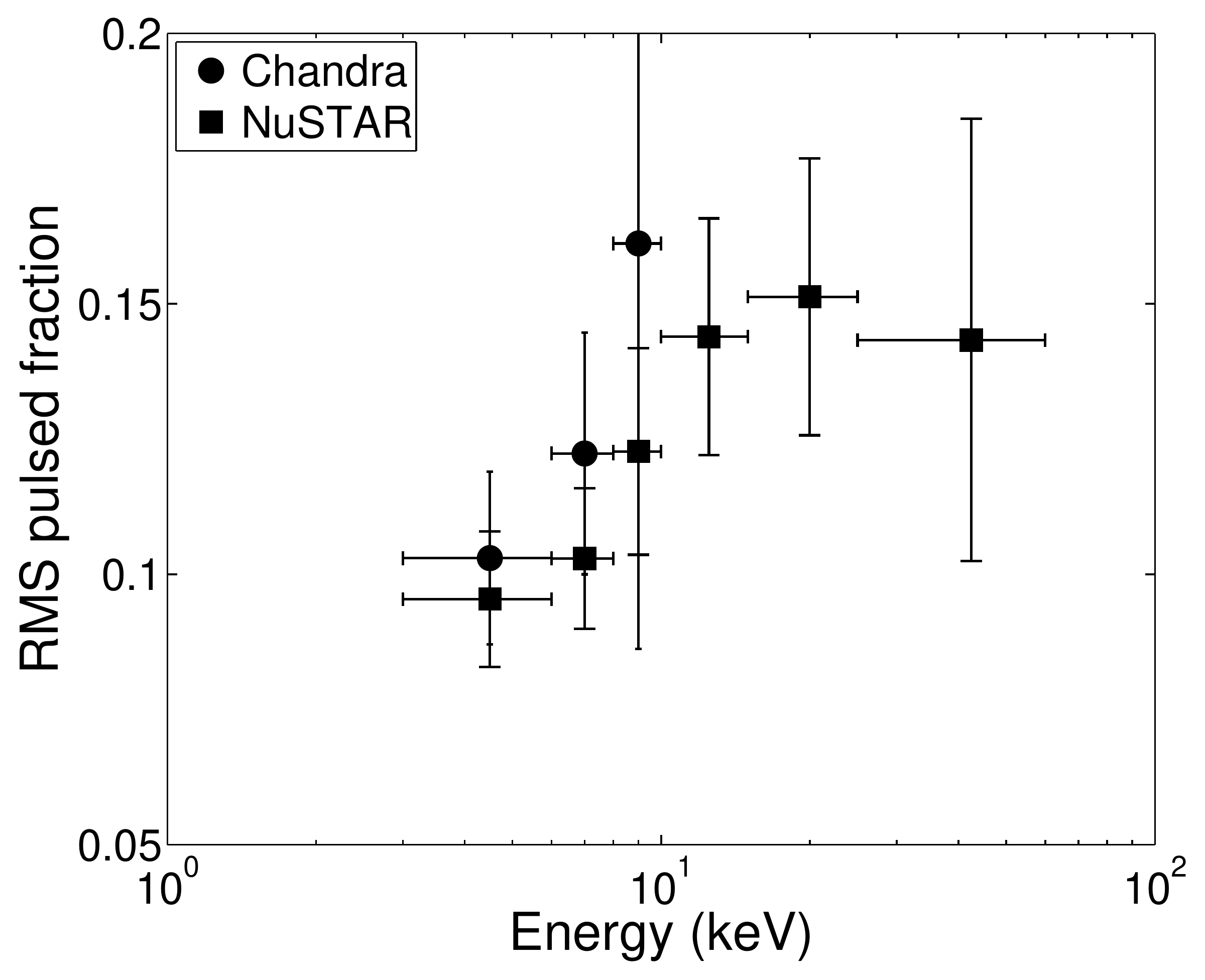}
\caption{Persistent emission pulse fraction as a function of energy for \nustar, squares, and \chandra, circles.}
\label{rmsVSene}
\end{center}
\end{figure}

We fit the different PPs with a sine plus cosine function
of the form \citep{bildsten97ApJ:PP},

\begin{equation}
\label{sinecosine}
C(\phi)=C_{\rm mean}+\sum_{\rm k=1}^{\rm m}[A_{\rm k} sin(2\pi {\rm k}\phi) + B_{\rm k}cos(2\pi {\rm k}\phi)]
\end{equation}

\noindent where $C(\phi)$ is the count rate at phase bin $\phi$,
$C_{\rm mean}$ is the average count rate throughout the PP,
and $A_{\rm  k}$, $B_{\rm k}$ are the coefficients of the different
harmonics $\rm k$ of the sine and cosine functions. The PP
is nearly sinusoidal. The second harmonic contribution is highest in
between 8 and 10~keV ($19\pm10\%$) for \chandra, and between 6 and
8~keV ($8\pm1\%$) for \nustar.

The rms pulsed fraction (PF) is defined as \citep{bildsten97ApJ:PP},

\begin{equation}
PF_{\rm rms}=\left[0.5 \sum_{\rm k=1}^{\rm m}(A_{\rm k}^{2}+B_{\rm
    k}^{2})-(\sigma_{A,\rm k}^{2}+\sigma_{B,\rm
    k}^{2})\right]^{0.5}/C_{\rm mean}
\end{equation}

\noindent where $\sigma_{A,\rm k}$ and $\sigma_{B,\rm k}$ are the 1
sigma standard deviations on the model coefficients.

We calculate the PF of the persistent emission PPs in
different energy bands and find a slight energy dependence in both
\chandra\ and \nustar\ data. At the lowest  energies, $3-6$~keV, the PF
is about $10\%$. It increases to  $15\%$ in the $10-15$~keV energy
range, and remains constant at higher energies
(Figure~\ref{rmsVSene}). We find the same dependence and PF values
during the dip intervals in both \chandra\ and \nustar.

We do not perform timing analysis on the burst intervals. Both
\chandra\ and \nustar\ suffer high instrumental dead-time during
bursts, and \chandra\ data suffer a small pile-up percentage at the peak
of the bursts ($10\%$), all of which distorts the burst PP.

\subsection{Spectral analysis}
\label{specana}

We perform our spectral analysis using XSPEC \citep{arnaud96conf}
version 12.8.1. The photo-electric cross-sections of \citet{
  verner96ApJ:crossSect} and the abundances of \citet{
  wilms00ApJ} are used throughout to account for absorption by neutral
gas. All quoted uncertainties are at the $1\sigma$ level, unless
otherwise noted.

\subsubsection{Persistent Emission}

\begin{figure}[t]
\begin{center}
\includegraphics[angle=0,width=0.5\textwidth]{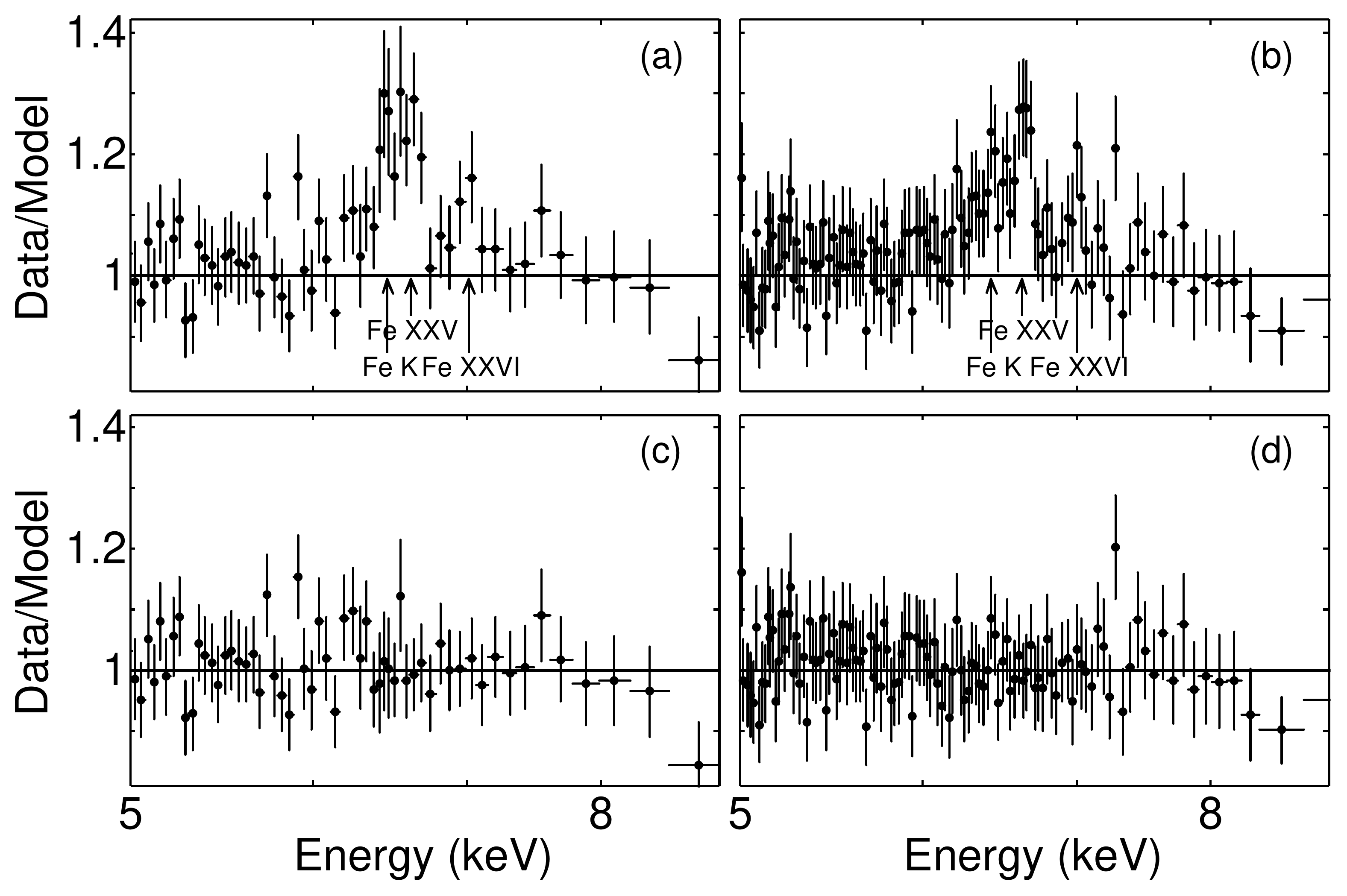}\\
\caption{{\sl Panel (a).} Data to model ratio of a PL fit to the
  \chandra\ persistent emission spectrum. {\sl Panel (b).} Data to
  model ratio of a PL fit to the \chandra\ dips+persistent emission
  spectrum. In both panels, we ignore the 6.0-8.0~keV range in the
  fits. {\sl Panel (c).} Data to model ratio of a PL and 3 Gaussian
  lines to the persistent emission spectrum. {\sl Panel (d).} Data to
  model ratio of a PL and 3 Gaussian lines to the dips+persistent
  emission spectrum. The arrows at the Gaussian centroid energies
  indicate their possible identification. Data are rebinned for
  clarity.}
\label{specPersisAIO}
\end{center}
\end{figure}

We extract the HEG~m1 spectrum of the persistent emission
intervals as defined in Section~\ref{timeana} and fit it (binned to a
S/N ratio of 10) with an absorbed PL. Residuals in the form of narrow
emission lines are present in the spectra in the energy range
$6.0-7.5$~keV (Figure~\ref{specPersisAIO}).

To properly model the continuum we ignore data in the energy range
$6.0-8.0$~keV. The absorbed PL provides a good fit to the data with a
reduced $\chi^2$ of 0.7 for 247 degrees of freedom (d.o.f.). We also
fit the continuum with a black-body function (bbody in XSPEC, BB
hereafter), and although we find a statistically acceptable fit with a
reduced $\chi^2$ of 0.9, significant fit residuals at low and high
energies ($<3$ and $>8$~keV) are present. A diskbb model gives as good
a fit as the PL with a reduced $\chi^2$ of 0.76; however, the
temperature $T_{\rm in}\approx6.0$~keV of the inner disk, is too high
for an accreting pulsar. We conclude that a simple PL is sufficient to
explain the \chandra\ data alone.

We add the 6.0 to 8.0~keV data, including three Gaussian lines at
6.4, 6.65, and 7.0 keV to account for the residuals, and re-fit. We
find a reduced $\chi^2$ of 0.7 for 291 d.o.f
(Figure~\ref{specPersisAIO}). Table~\ref{persisProp} lists the best
fit parameters to the  persistent emission continuum and features,
along with their $1\sigma$ uncertainties. The addition of the 6.4,
6.65, and 7.0~keV Gaussian components (one at a time) improve the fit
by a $\Delta\chi^2$ of 24, 31, and 10, respectively. We find that all
three lines are narrow with comparable fluxes and Equivalent Widths
(EW) within uncertainties. Most likely the lines are due to neutral or
near-neutral Fe, and highly ionized Fe XXV (He-like) and Fe XXVI
(H-like). We note here that there is a CCD gap in the HEG~m1
at 6.3~keV, right below the energy of the 6.4~keV Fe feature,
resulting in loss of counts at that energy, which renders the fit
parameters of this line uncertain.

%-----------
% Table  3
%-----------
\begin{table}[!t]
\caption{\chandra\ HEG~m1 best fit parameters for the dip, persistent, and dip+persistent emission intervals.}
\label{persisProp}
\newcommand\T{\rule{0pt}{2.6ex}}
\newcommand\B{\rule[-1.2ex]{0pt}{0pt}}
\begin{center}{
\resizebox{0.5\textwidth}{!}{
\begin{tabular}{l c c c}
\hline
\hline
\T\B & Persistent & Dips & Dips+Persistent \\
\hline
$N_{\rm H} (10^{22}$~cm$^{-2)}$\T\B                        & $8.8\pm0.3$ & $9.0_{-0.4}^{+0.5}$  & $8.9\pm0.2$ \\
\hline
$\Gamma$\T\B                                                       & $1.16_{-0.05}^{+0.03}$& $1.2\pm0.1$  &  $1.17_{-0.04}^{+0.03}$ \\
Norm.$^{a}$  \T\B &$1.01_{-0.08}^{+0.07}$& $0.95_{-0.1}^{+0.2}$ & $0.96_{-0.05}^{+0.04}$\\
$F_{\rm PL}$ ($10^{-8}$)$^{b}$   \T\B & $1.22_{-0.03}^{+0.02}$& $1.05_{-0.03}^{+0.02}$  & $1.15\pm0.02$ \\
\hline
$E_1$ (keV)                                                           \T\B &$6.45_{-0.03}^{+0.06}$&$6.4_{-0.2}^{+0.1}$ & $6.44\pm0.06$ \\
$\sigma_1$ (eV)                                                         \T\B & $45_{-17}^{+95}$&$700_{-400}^{+300}$  & $250_{-70}^{+90}$ \\
$EW_1$ (eV)                                                               \T\B & $31\pm14$ & $220_{-100}^{+140}$ & $81\pm21$ \\
$F_1$ ($10^{-11}$)$^{b}$       \T\B  & $4.0_{-1.3}^{+5.0}$  & $17_{-12}^{+10}$ & $9.0\pm2.0$ \\
\hline
$E_2$ (keV)                                                         \T\B  & $6.63_{-0.02}^{+0.03}$& $6.66_{-0.04}^{+0.03}$  & $6.65_{-0.02}^{+0.01}$ \\
$\sigma_2$ (eV)                                       \T\B  & $55_{-25}^{+21}$ & $42_{-26}^{+42}$ & $33_{-13}^{+15}$ \\
$EW_2$ (eV)                                                                \T\B & $42_{-15}^{+17}$ & $17_{-10}^{+12}$ & $17_{-6}^{+8}$ \\
$F_2$ ($10^{-11}$)$^{b}$        \T\B  & $5.3_{-2.3}^{+1.4}$ & $1.9_{-0.7}^{+1.4}$ &  $2.2\pm0.8$ \\
\hline
$E_3$ (keV)                                                         \T\B  & $7.00\pm0.03$& $6.98_{-0.04}^{+0.03}$  & $6.99\pm0.02$\\
$\sigma_3$ (eV)                                       \T\B&  $68_{-34}^{+68}$  & $11_{-11}^{+39}$ & $32_{-16}^{+30}$ \\
$EW_3$ (eV)                                                                \T\B & $25_{-13}^{+11}$ & $10_{-10}^{+12}$ & $15_{-7}^{+8}$ \\
$F_3$ ($10^{-11}$)$^{b}$         \T\B & $3.0_{-1.4}^{+1.7}$& $0.9_{-0.7}^{+0.9}$ & $1.7_{-0.5}^{+0.9}$ \\
\hline
$F_{\rm tot}$ ($10^{-8}$)$^{bc}$   \T\B & $1.23\pm0.02$ & $1.08_{-0.03}^{+0.02}$  & $1.16\pm0.02$ \\
\hline
\end{tabular}}}
\end{center}
\begin{list}{}{}
\item[{\bf Notes.}] $^a$ In units of photons~keV$^{-1}$
  cm$^{-2}$~s$^{-1}$. $^b$ In units of erg~cm$^{-2}$~s$^{-1}$. $^c$
  Total unabsorbed flux. $F_{\rm PL}$ and $F_{\rm tot}$ are caclualted
  in the 0.5-70~keV energy range. Fluxes of the Gaussian components
  are in the 6-8~keV range.
\end{list}
\end{table}
%-----------
% Table  3
%-----------

\begin{figure*}[t]
\begin{center}
\includegraphics[angle=0,width=0.8\textwidth]{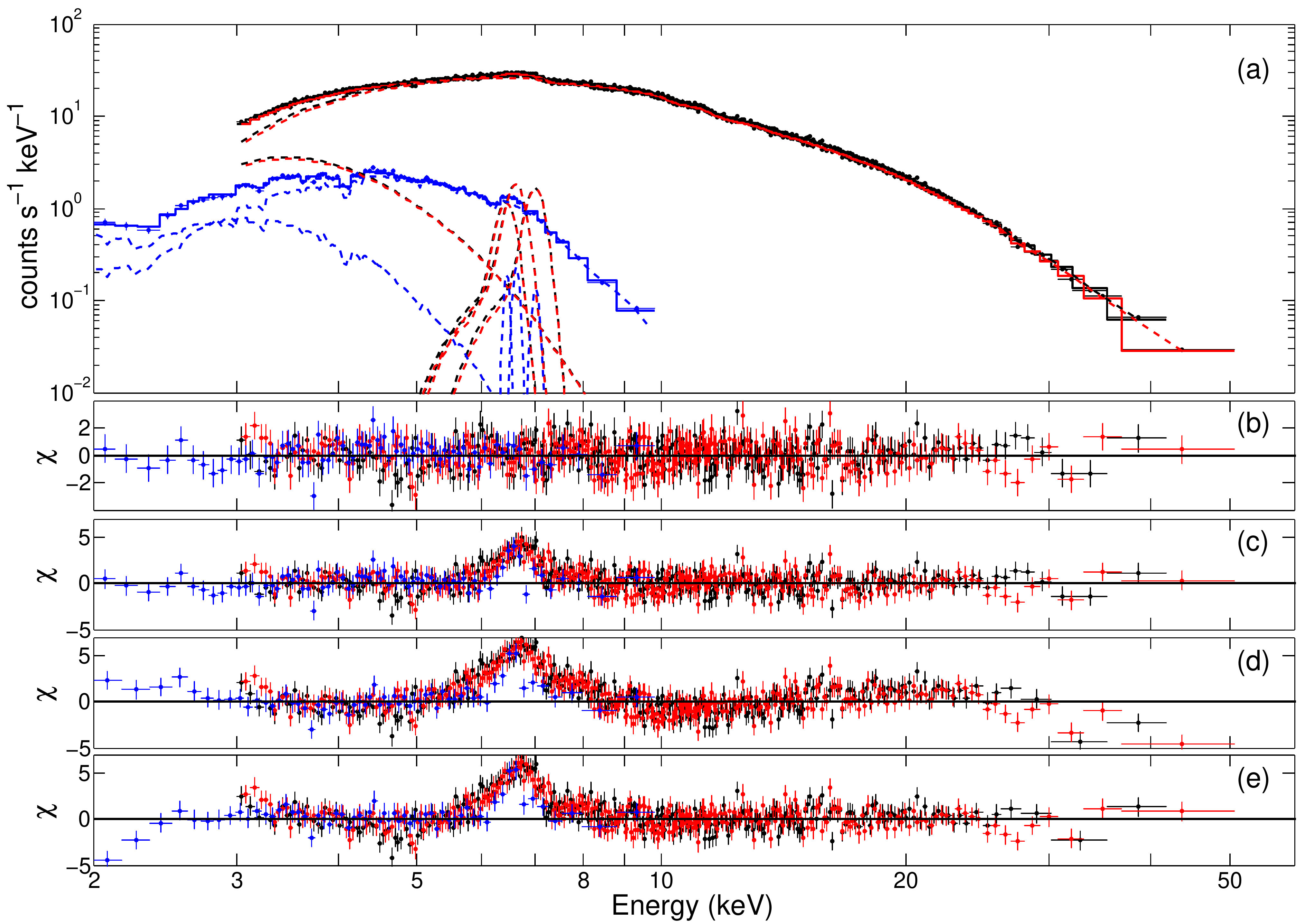}\\
\caption{{\sl Panel a.} Best fit model to the {\it NuSTAR} and \chandra/HEG~m1
  spectra of the persistent emission interval. The model consists of a
  BB, 3 Gaussians, a 10~keV feature modeled as {\sl cyclabs}, and a
  {\sl cutoffPL}, all modified by absorption. A constant normalization
  is also included for instrument cross-calibration
  uncertainties. Dashed lines represent the different additive
  components. {\sl Panel b.} Residuals of the data from the best-fit
  model. {\sl Panel c.} Ignoring the 6.0-8.0 keV data and excluding
  the 3 Gaussian lines. {\sl Panel d.} Excluding the 10~keV feature,
  and fixing $N_{\rm H}$ to the \chandra\ value. {\sl Panel e.}
  Letting $N_{\rm H}$ free to vary. Data have been refit in panels
  {\sl c}, {\sl d}, and {\sl e}. In all 5 panels, the black, red, and
  blue points are the \nustar\ module A, module B, and the \chandra\
  HEG~m1 data, respectively. \chandra\ data in the bottom two panels
  are binned-up for clarity. See text for more details.}
\label{nusChanFit}
\end{center}
\end{figure*}

Next, we extract the \nustar\ FPMA and FPMB spectra, in the 3$-$70
keV range for the persistent intervals simultaneous with \chandra, and
group them to have a S/N ratio of 25. We fit the \nustar\ spectra
and the HEG~m1 spectrum simultaneously, including a normalization
factor to all model fits to take into account any cross-calibration
uncertainties between the three instruments. We link all fit
parameters between the three spectra except for the normalization
factor. Both \nustar\ spectra show a broad emission line centered
at around 6.65~keV, most likely corresponding to the Fe line complex
detected with \chandra. Hence, for our initial fits, we exclude from
both \chandra\ and \nustar\ the $6.0-8.0$~keV energy range. A single
component model, i.e., PL or BB, does not give a satisfactory fit with
a reduced $\chi^2>2$. An absorbed {\sl cutoffPL} (a PL with an
exponential rolloff) model gives a better fit, but it is also
statistically unacceptable (reduced $\chi^2=1.8$). The addition of a
BB model improves the fit dramatically with a reduced
$\chi^2=1.3$ (Figure~\ref{nusChanFit}, panel~e). However, this model
results in an absorption hydrogen column density three times lower
than what we get from \chandra\ alone, resulting in large residuals at
the lower end of the spectrum, along with residuals at 10~keV. Hence,
we fix $N_{\rm H}$ to the \chandra\ value of
$9.0\times10^{22}$~cm$^{-2}$. This gives a similar fit quality
compared to the above, but emphasizes the residuals around 10~keV in
the form of a broad trough (Figure~\ref{nusChanFit}, panel~d). Adding
a negative broad feature ({\sl cyclabs}\footnote{We tried three
  different negative broad components to model the 10~keV feature,
  {\sl cyclabs}, {\sl gabs}, and an additive Gaussian component with
  negative normalization. {\sl cyclabs} gives a slightly better fit
  than the other two with $\Delta\chi^2=24$ for the same number of
    d.o.f. Hence, we use {\sl cyclabs} in the rest of the analysis.}
  in XSPEC) to the model and thawing $N_{\rm H}$ results in the best fit
  to the continuum with a reduced $\chi^2$ of 0.95 (for 767 d.o.f.).

We also investigate the effect of other model continua on the
presence and shape of the 10~keV feature. First, instead of a {\sl
  cutoffpl} model, we fit the spectrum using (1) a Fermi-Dirac form
of cutoff \citep[{\sl fdcut},][]{tanaka86LNP:fdcut}, and (2) a
negative$-$positive PL exponential \citep[{\sl
  npex},][]{mihara95PhDT:npex}, both used for fitting accreting X-ray
pulsar spectra. Neither model, modified by absorption, gives a
  good fit to the data (reduced $\chi^2\approx1.9$). Adding a BB
  results in a reduced $\chi^2$ of 1.3. We then remove the BB
  component and add a negative broad feature to the models. Both {\sl
    fdcut} and {\sl npex} give a reduced $\chi^2$ of 1.1 but fail to
  reproduce the soft part of the spectrum. Adding both a BB component 
and a negative feature gives a good fit to the data in both cases
(reduced $\chi^2$ of 0.94 and 0.95 for 765 and 766 d.o.f for {\sl
  npex} and {\sl fdcut}, respectively). We conclude that the 10~keV
feature and the BB component are present in the data regardless of the
shape of the continuum used. We, therefore, adopt the simplest
empirical model, i.e., {\sl cutoffPL}, for the rest of our analysis
since it has less free parameters than the above two for comparable
fit results. Moreover, the parameters of the BB component and the 10
keV feature are consistent within 1$\sigma$ in all three models.

Finally, using the {\sl cutoffPL} continuum model, we add the
6.0$-$8.0~keV data. We include three Gaussian lines with centroid
energies and widths fixed to the values derived from the \chandra\
data (reduced $\chi^2$ of 0.91 for 919 d.o.f.). We also fit one
Gaussian line to the data, with all parameters left free to vary. We
find a line centroid energy $E=6.69\pm0.03$~keV and a width
$\sigma=0.42_{-0.04}^{+0.06}$~keV (reduced $\chi^2$ of 0.92 for
  919 d.o.f.).

We conclude that our best fit model for the \nustar\ and HEG~m1
persistent emission spectra consists of a BB, a {\sl cutoffPL}, a
$10$~keV feature, three Gaussian lines with centroid energies and
widths fixed to the \chandra-alone values (or 1 Gaussian component
with all parameters left free to vary), all affected by neutral
absorption ({\sl tbabs} in XSPEC), and a constant
normalization. Table~\ref{NusChanPersisProp} gives the best fit
parameters.

%-----------
% Table  4
%-----------
\begin{table}[!t]
\caption{Best fit parameters for the dip, persistent,
  dip+persistent, and burst emission intervals for the combined \nustar\ $+$\chandra\ data.}
\label{NusChanPersisProp}
\newcommand\T{\rule{0pt}{2.6ex}}
\newcommand\B{\rule[-1.2ex]{0pt}{0pt}}
\begin{center}{
\resizebox{0.5\textwidth}{!}{
\begin{tabular}{l c c c c}
\hline
\hline
\T\B & Persistent & Dips  & Dips+Persistent & Bursts \\
\hline
$N_{\rm H} (10^{22}$~cm$^{-2})$\T\B                            & $11.2\pm0.7$& $9.7\pm1.0$  & $10.6\pm0.6$ & $7.3\pm0.7$\\
\hline
$kT_{\rm BB}$ \T\B                                                        & $0.55\pm0.03$ & $0.52\pm0.03$& $0.52\pm0.02$ & $---$ \\
Norm.$_{\rm BB}$ ($10^{-2}$)$^{a}$         \T\B               & $4.0_{-0.8}^{+1.0}$ & $4.6_{-1.2}^{+1.5}$& $4.1_{-0.7}^{+0.8}$ & $---$ \\
$F_{\rm BB}$ ($10^{-9}$)$^{b}$       \T\B                        & $4.5\pm0.6$& $4.3\pm0.8$  & $4.5_{-0.5}^{+0.4}$ & $---$ \\
\hline
$E_1^d$ (keV)                            \T\B                               & $6.45$ & $6.4$ & $6.44$ & $---$ \\
$\sigma_1^d$ (eV)               \T\B                                           & $45$& $700$  & $250$ & $---$ \\
EW$_1$ (eV)                         \T\B                                        & $22_{-5}^{+8}$& $141_{-33}^{+36}$  & $47_{-7}^{+5}$ & $---$ \\
$F_1$ ($10^{-11}$)$^{b}$     \T\B                                   & $2.9\pm0.8$ &   $12\pm2$    & $5.5\pm0.7$ & $---$ \\
\hline
$E_2^d$ (keV)              \T\B                                             & $6.63$& $6.66$  & $6.65$ & $---$ \\
$\sigma_2^d$ (eV)     \T\B                                    & $55$& $42$  & $33$ & $---$ \\
EW$_2$ (eV)              \T\B                                                   & $33_{-5}^{+7}$& $27_{-4}^{+6}$  & $22\pm3$ & $---$ \\
$F_2$ ($10^{-11}$)$^{b}$    \T\B                                  & $4.5_{-0.6}^{+0.5}$& $3.2_{-0.5}^{+0.6}$  &  $3.2\pm0.5$ & $---$ \\
\hline
$E_3^d$ (keV)              \T\B                                             & $7.01$& $6.98$  & $6.99$& $---$ \\
$\sigma_3^d$ (eV)   \T\B                                     &  $68$& $11$  & $32$ & $---$ \\
EW$_3$ (eV)                          \T\B                                       & $36_{-4}^{+6}$& $15_{-5}^{+4}$  & $27_{-3}^{+2}$ & $---$ \\
$F_3$ ($10^{-11}$)$^{b}$    \T\B                                   & $4.5_{-0.4}^{+0.5}$& $1.7\pm0.4$ & $3.6\pm0.3$ & $---$ \\
\hline
$E_b$ (keV)$^c$              \T\B                                   & $6.69\pm0.03$  & $6.71\pm0.04$ & $6.70_{-0.03}^{+0.02}$& $---$ \\
$\sigma_b$ (eV)   \T\B                                    &  $400_{-40}^{+60}$ & $420\pm50$  & $420_{-20}^{+40}$ & $---$ \\
EW$_b$ (eV)                          \T\B                               & $123_{-12}^{+16}$ & $133_{-13}^{+9}$ & $129_{-9}^{+10}$ & $---$ \\
$F_b$ ($10^{-11}$)$^{b}$    \T\B                                   & $15_{-1}^{+2}$& $14_{-1}^{+2}$ & $15\pm1$ & $---$ \\
\hline
$E_{\rm 10~keV}$ (keV)                \T\B                            & $9.9\pm0.2$& $9.5_{-0.4}^{+0.3}$ & $9.9\pm0.1$ & $10.5\pm0.3$\\
$\sigma_{\rm 10~keV}$ (keV)            \T\B                             &  $3.7\pm_{-0.3}^{+0.4}$  & $4.2_{-0.5}^{+0.6}$& $3.6_{-0.2}^{+0.3}$ & $0.8_{-0.3}^{+0.8}$ \\
$d_{\rm 10~keV}$             \T\B                                       &  $0.16\pm0.02$ & $0.18\pm0.02$ & $0.17\pm0.01$ & $0.15\pm0.04$ \\
\hline
$\Gamma$                   \T\B                           & $0.0_{-0.1}^{+0.2}$ & $0.0\pm0.1$  &  $0.00\pm0.04$ & $0.2\pm0.1$\\
$E_{\rm fold}$                  \T\B                                        & $7.0\pm0.2$  & $7.0\pm0.2$& $7.1\pm0.1$ & $7.6\pm0.5$\\
Norm$^{a}$       \T\B                                       & $0.29_{-0.02}^{+0.03}$& $0.26\pm0.03$ & $0.30\pm0.02$ & $1.9_{-0.2}^{+0.3}$\\
$F_{\rm PL}$ ($10^{-8}$)$^{b}$     \T\B                            & $2.49\pm0.04$  & $2.24_{-0.05}^{+0.06}$& $2.37\pm0.03$ & $9.8\pm0.2$\\
\hline
$F_{\rm tot}$ ($10^{-8}$)$^{be}$     \T\B                            & $2.62\pm0.02$  & $2.30_{-0.02}^{+0.03}$& $2.47\pm0.01$ & $8.9^{+0.2}_{-0.1}$\\
\hline
\end{tabular}}}
\end{center}
\begin{list}{}{}
\item[{\bf Notes.}] $^a$ In units of photons~keV$^{-1}$
  cm$^{-2}$~s$^{-1}$. $^b$ In units of erg~cm$^{-2}$~s$^{-1}$. $^{c}$
  Fitting one Gaussian line to the Fe line complex. $^{d}$ Fixed to
  the \chandra\ best fit results. $^e$ Total unabsorbed flux. $F_{\rm
    BB}$, $F_{\rm PL}$, and $F_{\rm tot}$ are caclualted in the
  0.5-70~keV energy range. Fluxes of the Gaussian components are in
  the 6-8~keV range.
\end{list}
\end{table}
%-----------
% Table  4
%-----------

\subsubsection{Dips}
\label{dipsSpectra}

We extract the \chandra\ HEG~m1 spectra for each of the dip intervals
separately (here we define the duration of a dip interval as starting
from the end of a burst until the time it recovers to 95\% of the
persistent level). We fit all 7 intervals simultaneously with an
absorbed PL model. The hydrogen column density is linked for all
spectra to ensure that they are all equally absorbed. First, we allow
the PL indices and normalizations to vary and we find that the PL index
is consistent across all spectra. Therefore, we also keep the indices
linked. We do not find any flux variability between the different
dips, and estimate an average flux and 1$\sigma$ standard deviation of
$(1.05\pm0.04)\times10^{-8}$~erg~s$^{-1}$~cm$^{-2}$. We also find no
variability in the fluence deficiency during the different dips (the
actual deficit of fluence from the persistent emission during a dip)
with an average of $(1.2\pm0.2)\times10^{-6}$~erg~cm$^{-2}$.
Table~\ref{dipProp} gives the spectral results for the individual dip
intervals.

Motivated by the differences in the count rates at the dip minima
(Figure~\ref{fitDips}), we extract the HEG~m1 spectra of all seven dips in an 80~s
interval centered at $t_{\rm min}$ and fit them simultaneously with an
absorbed PL model, keeping only the normalizations free. We do not find
any flux variability (at the $>3\sigma$ level) in the minimum level
the dips reach after each burst (Table~\ref{dipProp}). We repeat the
analysis for a 40~s interval centered at $t_{\rm min}$, and
reach the same conclusions.

Next, we extract the HEG~m1 spectrum for dip intervals {\it
  collectively}, and group them so that each bin has a S/N ratio of
10. We fit the spectrum with an absorbed PL and find an absorption
hydrogen column density and a PL index consistent within errors with
the results we find for the persistent emission intervals
(Table~\ref{persisProp}). We find an unabsorbed average flux
$F=1.05_{-0.03}^{+0.02}\times10^{-8}$~erg~s$^{-1}$~cm$^{-2}$. The
ratio of the average flux (over the entire dip intervals) to the
persistent emission flux is $0.86$, or a $14$\% drop in the persistent
emission flux.

Finally, we find that some residuals are present in the HEG~m1
at high energies ($>6$~keV), in the
form of excess emission, similar to what is seen in the persistent
emission spectrum. Hence, we add three Gaussian lines at 6.4, 6.65,
and 7.0 keV. Table~\ref{persisProp} lists the best fit parameters to
the dip intervals. Unfortunately, due to the low number of counts in the
dip spectra, we are not able to adequately constrain the parameters of
the Gaussian components, except for their energies. Nonetheless,
comparing the dip and persistent emission fit parameters, we do not
find dramatic changes in the line properties. These results are
discussed in Section~\ref{discuss}.

We then apply the same model we used to fit the persistent interval
\chandra+\nustar\ broadband spectrum to the dip broadband spectrum. We
find very similar values between persistent and dip intervals, except
for the flux of the {\sl cutoffPL} component, which decreased by 10\%
during dips. The fluxes of the BB and the Fe line component
(considering one Gaussian line fit to the 6.7~keV excess) did not
change within their $1\sigma$ error. These results are
shown in Table~\ref{NusChanPersisProp}.

\subsubsection{Persistent Emission and Dips}

To achieve better S/N ratio for the spectral fitting of the lines, we
extract the HEG~m1 spectrum of the persistent and dip intervals
together. We fit the spectrum with an absorbed PL model and three
Gaussian emission lines (Figure~\ref{specPersisAIO}).
Table~\ref{persisProp} lists the best fit parameters along with their
$1\sigma$ uncertainties.

According to the persistent+dip spectrum, which has better statistics
than the persistent or dip spectra alone, the highly ionized lines are
narrow with widths of about 30~eV. They also contribute
similarly to the total flux. The neutral Fe, on the other hand, has a
larger width and flux compared to the other two.

The broadband model used to fit the persistent and dip spectra alone
is also successfully fit to the dips+persistent \nustar\ and
\chandra\ spectra (Table~\ref{NusChanPersisProp}).  All the parameters
of the best-fit model were compatible with the persistent and dip
fits. A one Gaussian emission line fit to the 6.7~keV excess results
in a width $\sigma=0.42$~keV and an EW=129~eV. Finally, we also fit
the 6.7~keV excess  using a diskline model \citep{ 
    fabian89MNRAS:diskline} and find $E=6.63_{-0.04}^{+0.05}$~keV,
  $R_{\rm in}=130_{-80}^{+240}$~GM/c$^2$
  ($3_{-2}^{+5}\times10^{7}$~cm), and $i>35$\degree\ (all quoted
  uncertainties are at the  $3\sigma$ level).

\subsubsection{Bursts}
\label{burstSpec}

We extract the HEG~1 spectrum for each of the seven bursts seen with
\chandra, and fit them simultaneously with an absorbed PL. We link the
hydrogen column density in the fit. We find a consistent PL index for
all spectra and, therefore, we also link the index thereafter. The PL
normalizations are left free to vary, to account for any flux
variability between the bursts. We find $N_{\rm
  H}=(9.0\pm1.0)\times10^{22}$~cm$^{-2}$, and a PL index of
$\Gamma=1.2\pm0.2$. We report in Table~\ref{burstProp} all burst
spectral parameters.

We find very similar energetics between the different bursts; flux
and luminosity variability (at the $3\sigma$ level) is observed only
between bursts \#4 and \#7 (these are the shortest and the longest
burst, respectively, Table~\ref{burstProp}). In terms of fluence, all
bursts emitted comparable (at the $2\sigma$ level) amounts of
energy. The mean and 1$\sigma$ standard deviation of the fluxes and
fluences are $(7.4\pm1.6)\times10^{-8}$~erg~cm$^{-2}$~s$^{-1}$, and
$(10.3\pm1.6)\times 10^{-7}$~erg~cm$^{-2}$, respectively. We also
convert the 1~s peak count rates of the seven bursts to fluxes using
PIMMS (due to the low count statistics). We find an average peak flux
of $2.0\times10^{-7}$~erg~s$^{-1}$~cm$^{-2}$,  equivalent to a
luminosity of $1.5\times10^{39}$~erg~s$^{-1}$ at 8~kpc. This value
should be regarded as a lower limit due to a $10\%$ pile-up, and
the narrow energy band for which it was derived ($0.5-10$~keV;
\nustar\ data were excluded from this analysis due to severe dead-time
effects).

We extract the HEG~1 spectrum of all seven bursts collectively,
grouped to a S/N of 7, to search for any features present in their
added spectrum. We fit the spectrum with an absorbed PL, and find a
hydrogen column density $N_{\rm
  H}=(1.0\pm0.1)\times10^{23}$~cm$^{-2}$, and a PL index
$\Gamma=1.2\pm0.1$. No prominent absorption and/or emission features
are seen in the spectrum.

\begin{figure}[th!]
\begin{center}
\includegraphics[angle=0,width=0.5\textwidth]{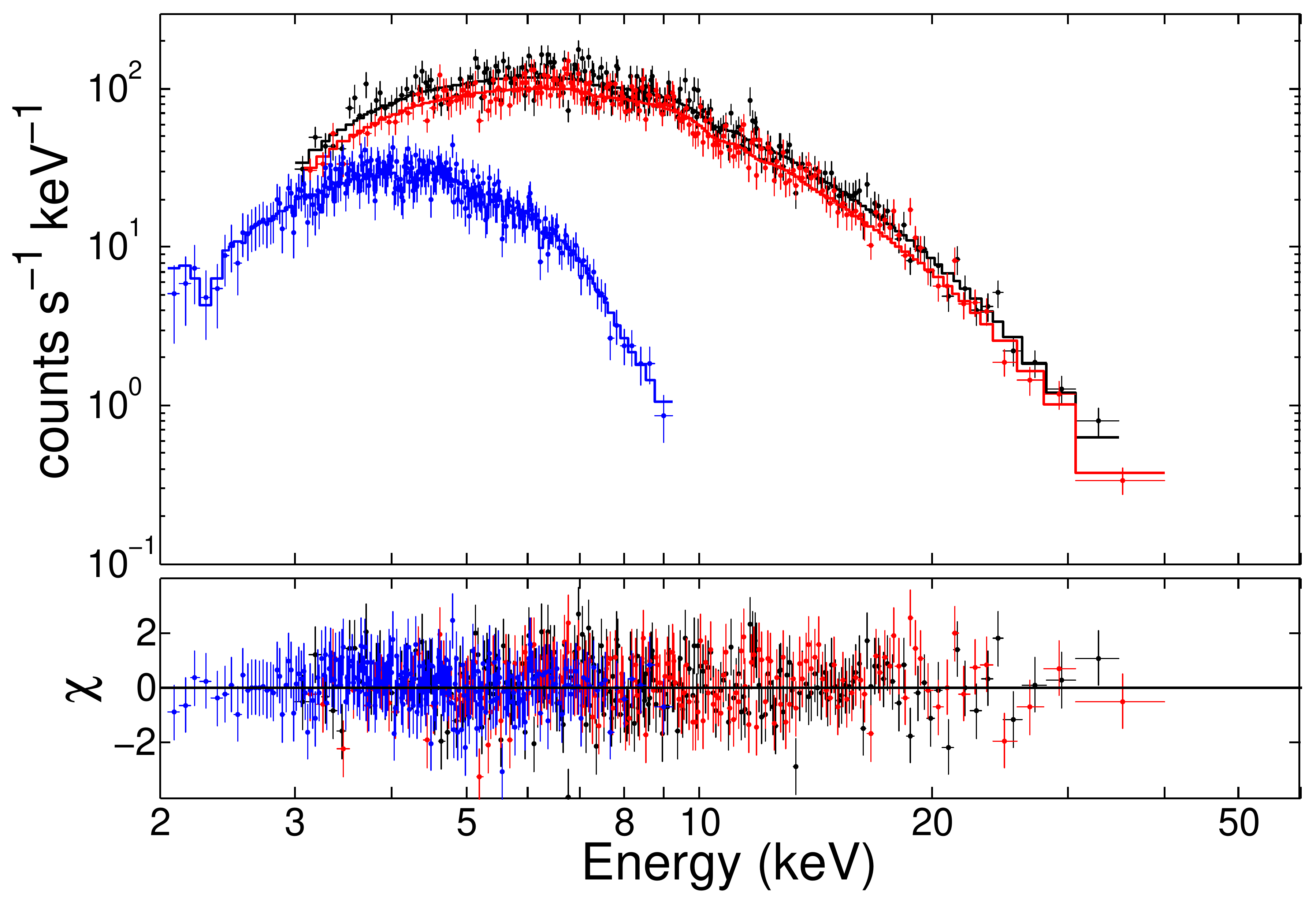}\\
\caption{{\sl Upper panel.} Data and best fit model to the
  \chandra+\nustar\ burst spectra. The model consists of a  {\sl
    cutoffPL} and a 10~keV feature, modified by absorption. {\sl Lower
    panel.} Deviations from the fit in terms of sigmas. Black, red, and
  blue points are the \nustar\ module A, module B, and the \chandra\
  HEG~m1 data, respectively.}
\label{specBurstAIO}
\end{center}
\end{figure}

We then extract the $3-70$~keV \nustar\ spectrum of the three bursts
seen simultaneously with \chandra, binned to a S/N ratio of
15. We fit the spectrum with an absorbed {\sl cutoffPL} resulting in a
reduced $\chi^2$ of 0.89 for 637 d.o.f.. Some residuals around 10~keV
can be seen, similar to what we find in the persistent and dip
spectra. Including a {\sl cyclabs} feature to the model improved the
fit slightly, resulting in a reduced $\chi^2$ of 0.86 for 634 d.o.f.,
which is a $\Delta\chi^2$ of 17 for 3 additional
parameters. Figure~\ref{specBurstAIO} shows the data and best fit
model, while the parameter values are listed in
Table~\ref{persisProp}. We see no excess emission between 6 and
8~keV. We derive a 3$\sigma$ upper limit of 119~eV on the EW of a
line with centroid energy at 6.7~keV. We also derive a
3$\sigma$ upper limit on the flux of a line at the same energy and a
width of 0.4~keV of $1.4\times10^{-10}$~erg~s$^{-1}$~cm$^{-2}$. Using
the co-added \chandra\ burst spectra, we derive a 30~eV $3\sigma$ upper limit on
the EW of the neutral Fe, and $16$~eV on the highly ionized
species. Assuming a width of 0.05~keV for the Fe~K and the highly
ionized species, we find a $3\sigma$ flux upper limits of the order of
$10^{-10}$~erg~s$^{-1}$~cm$^{-2}$ for all three lines. These upper
limits indicate that if any of the lines we detect in the dip and
persistent spectra brightened proportionally to the burst flux (on the
average by a factor of 5), we should have been able to detect
them. However, if the line fluxes remained constant during the bursts,
their presence could be masked by the much brighter burst continuum.

Finally, we note that the BB component is not required
by the fit to the bursts spectrum at a high significance. However, the
hydrogen column density is lower at the $3\sigma$ level than the value
derived for the  dips and persistent intervals
(Table~\ref{NusChanPersisProp}). Hence, fixing  the column density at 
$10.6\times10^{22}$~cm$^{-2}$, we find residuals at the lower end of the
spectrum, which are well fit with a BB component with
$kT\approx0.6$~keV and a $0.5-70$~keV flux of
$(3\pm2)\times10^{-9}$~erg~s$^{-1}$~cm$^{-2}$, consistent with the BB
temperatures and fluxes during the persistent and dip intervals. These
results are discussed in Section~\ref{discussBBs}.

\subsection{Phase resolved spectroscopy}

\begin{figure}[th!]
\begin{center}
\includegraphics[angle=0,width=0.5\textwidth]{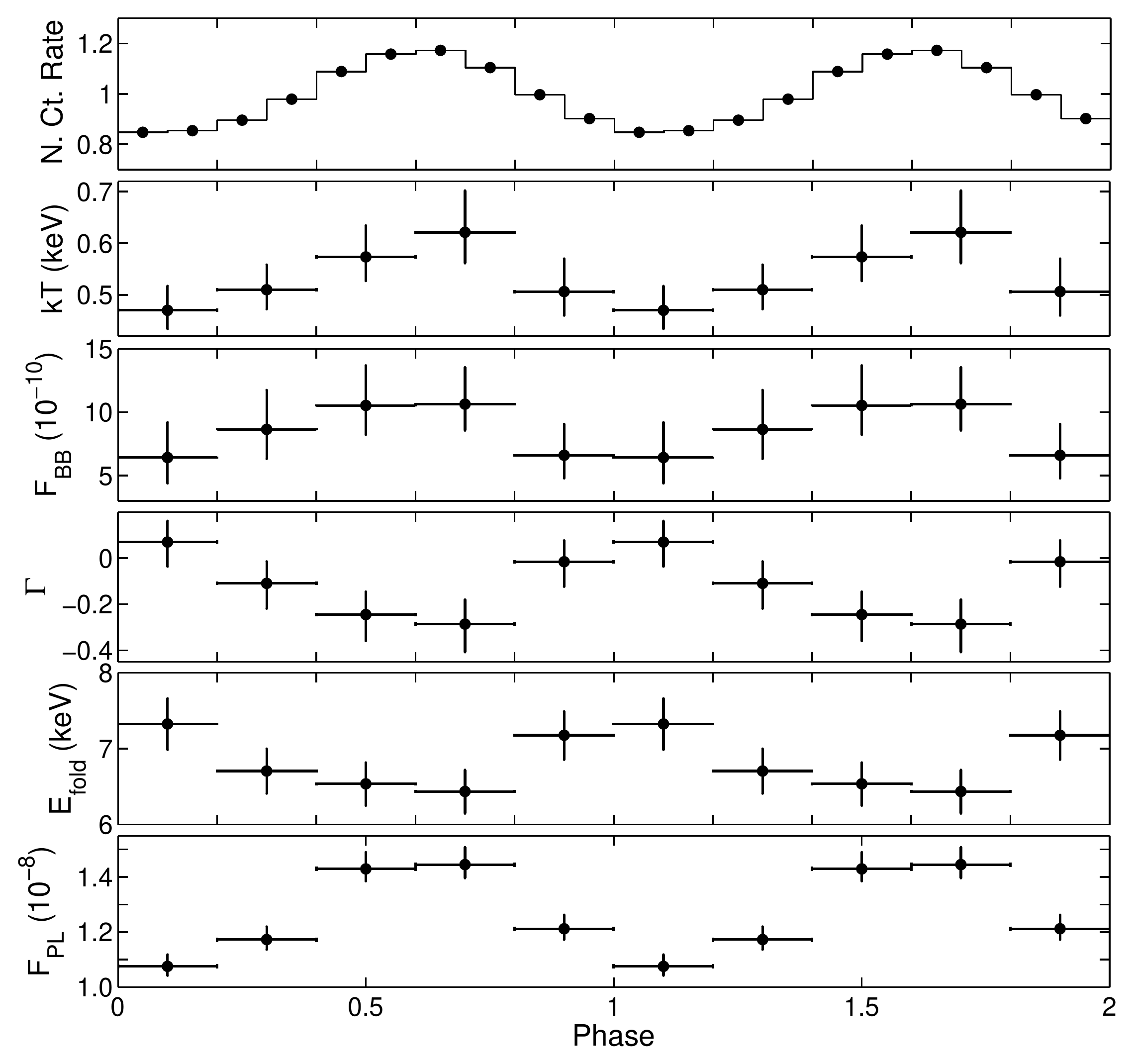}\\
\caption{Broad-band phase resolved spectroscopy during the persistent
  emission of the BP. From top to bottom, PP in the 3-70 keV range, BB
  temperature, BB flux, {\sl cutoffPL} index, energy roll-off, and
  {\sl cutoffPL} flux. See text for more details.} 
\label{phaseResSpec}
\end{center}
\end{figure}

We divide the broadband persistent emission spectrum into five pulse
phase bins, which we fit simultaneously with our best fit model
described above. We fit the $6.7$~keV excess energy with one Gaussian
line. We first leave all model parameters free to vary. We link one
model parameter after another (starting from the least variable
according to a $\chi^2$ test), and record the F-test significance at
each step to assess the significance of leaving the parameter free in
the fit. We find that the fit parameters of the 6.7 and 10~keV
features do not show significant changes with pulse phase. On the
other hand, the BB and {\sl cutoffPL} fit parameters tightly follow
the PP shape (Figure~\ref{phaseResSpec}). We find a decrease in the
photon index $\Gamma$ and an increase in the BB temperature indicating
that the X-ray spectrum hardens at pulse maximum. The rolloff energy
is also anti-correlated with the pulse shape.

\section{Discussion}
\label{discuss}

\subsection{Burst and dip origin}

Our temporal analysis shows that six out of the seven detected bursts
have comparable durations with an average of about $12$~s. These
consist of a single pulse with a faster rise than decay time. The
seventh burst detected by \chandra\ is the only outlier with a
duration of about 25~s, consisting of two pulses. These temporal
properties are similar to the properties of hundreds of bursts
recorded during the first two outbursts from the BP \citep[e.g.,
][]{woods99ApJ:1744}. 

From the \chandra\ data, we can derive the $\alpha$ parameter, the
ratio of the fluence in the persistent emission to the fluence in the
bursts. We find $\alpha<15$ for all bursts (except for the sixth one
where $\alpha=30$), with an average value of 10. The same value was also
derived during the first two outbursts from the source
\citep{mejia02ApJ:1744}. As pointed out by \citet{lewin96ApJ:1744},
this small value of $\alpha$ is inconsistent with thermonuclear
burning as the origin for the BP bursts. This value is consistent with
the observed bursts being type~II bursts, similar to what is seen in
the Rapid Burster, caused by some sort of instability associated with
the accretion disk \citep{lamb77ApJ:accDisk,baan79ApJ:accDisk,
  taam84ApJ:accDisk,lasota91:accDisk,spruit93ApJ:accDisk,
  cannizzo96ApJ:1744,dangelo10MNRAS:accDisk}. Unlike the Rapid
Burster, however, the BP does not display any correlation between the
fluence in a burst and the time to the following burst \citep[e.g.,
][]{kouveliotou96:1744}. This is also evidenced by our data, where the
fluence emitted during all seven bursts is constant while the
intervals between bursts changed by up to a factor of 4.

Similar to the previous two outbursts, the bursts we detect in
\chandra\ and \nustar\ are followed by a dip, where the
X-ray emission decreases by 10\% on average, and 40\% at
dip minimum. The emission exponentially recovers back to the pre-burst
persistent level, on time-scales of a few hundred seconds. We find
that the fluence in a burst and the integrated flux deficiency of the
following dip are consistent within $1\sigma$ (Tables~\ref{burstProp}
and \ref{dipProp}). We estimate an average burst fluence ($2-10$~keV)
of $(1.0 \pm 0.2)\times 10^{-5}$ erg cm$^{-2}$, and an average {\it
  missing} dip fluence of $(1.2\pm 0.2)\times 10^{-5}$~erg~cm$^{-2}$.
Such a correlation was also seen during the previous two outbursts
\citep[e.g.,][]{nishiuchi99ApJ:1744}. These authors suggested that the
energy emitted during a burst could be compensated by the deficit in
energy during the following dip. A very simple picture would be that
accretion-disk instabilities would allow for a sudden and rapid
increase of the mass-inflow rate onto the polar cap of the neutron
star from a reservoir (e.g., the accretion disk). The dips, then,
would be the result of a small fraction of the continuously accreted
matter disappearing to replenish this reservoir.

\subsection{X-ray emission properties}
\label{discussBBs}

The broadband spectrum of the BP is the typical spectrum of an
accreting X-ray pulsar at high accretion rates \citep[see e.g.,][for a
review]{coburn02ApJ:cycl}. It is well fit with a hard component,
modeled as a PL with an exponential rolloff, an Fe line complex, a
soft component modeled with a BB, and a 10~keV feature, all
modified by absorption. In the following, we will discuss these
different components and their interplay between persistent, dip, and
burst emission, except for the Fe line complex which is discussed in
Section~\ref{disFe}.

The high persistent X-ray luminosity of \src\ during the present
observation ($L_{\rm X}=1.9\times10^{38}$~erg~s$^{-1}$) implies that
the emission is coming from an accretion column, where the kinetic
energy of the infalling gas onto the polar cap is converted to
radiation via a radiative shock above the thermal mound \citep{
  basko75:xraypulsar}. Thermal photons from the mound, as well as
cyclotron and bremsstrahlung radiation, are converted to high-energy
photons via inverse Compton scattering. Hence, the resulting X-ray
spectrum will depend on several parameters such as the geometry of the
system and the properties of the compact source $-$ mainly its dipole
magnetic field $-$ among others \citep{becker07ApJ:xraypulsar}.  Even
with the small dipole magnetic field of \src, and its complicated
accretion geometry \citep{miller96ApJ:1744}, the parameters we derive
from our phenomenological fit compare reasonably well to other
accreting X-ray  pulsars \citep[e.g.,
][]{decesar13ApJ:1626,suchy11ApJ:1118,muller13:0115,furst14ApJ:1947}. The
photon index of the {\sl cutoffPL} is slightly lower than in most
cases implying a harder spectrum, which could be the result of the
higher luminosity of the source. Simply put, a higher accretion rate
onto the poles, would lead to a higher electron density in the
accretion column and to higher Compton $y-$parameter causing a harder
spectrum. The \src\ spectrum also shows a lower energy rolloff
compared to other sources, which could be due to the relatively low
magnetic field of the source \citep{coburn02ApJ:cycl}.

The 10~keV feature is not unique to \src, and has previously been
reported in other accreting X-ray pulsars, e.g., Vela~X-1 \citep[][see
also \citealt{coburn02ApJ:cycl} for a review]{furst13ApJ:herX1,
  furst14ApJ:velax1,muller12:1946}. This feature is not always
necessarily observed as an absorption trough and sometimes manifests
itself as a broad emission feature or a wiggle. It is believed to be
the result of modeling accreting X-ray pulsar spectra with simple
empirical functions, when the true physics giving rise to their X-ray
spectra is far more complicated, especially when they are emitting
near the Eddington limit \citep[see][for a discussion]{
  coburn02ApJ:cycl}. In a  few cases, however, such as in the case of
the Be/X-ray binary Swift~J1626.6$-$5156 \citep{decesar13ApJ:1626}, an
absorption line at 10~keV was interpreted as a cyclotron resonance
scattering feature (CRSF), evidenced by the presence of a weak second
harmonic and the fact that the B field strength, derived from the
line energy, was consistent to the value derived from the spin-up rate
of the source \citep{icdem11MNRAS:1626}. In the 10~ks of \nustar\ data
that we consider here, we find no evidence of a second harmonic at about
twice the energy of the 10~keV feature (i.e., $20$~keV), and the
$B-$field strength corresponding to the line energy
($B\approx9\times10^{11}$~G for a 10~keV line energy) is significantly
larger than the estimates we derive in Section 4.4. Moreover, CRSFs
usually show strong dependence with pulse phase \citep[e.g.,
][]{furst14ApJ:1947}, which we do not observe in our phase-resolved
spectroscopy. Hence, we consider the 10~keV
feature in \src\ spectrum to be a defect of our continuum
modeling. We note, however, that unlike the BP, other X-ray pulsars
showing 10~keV features invariably show CRSFs.

Soft excess emission is often modeled with a BB component in accreting
X-ray pulsars. \citet[][see also
\citealt{Ballantyne12ApJ:BBXRP}]{hickox04ApJ:bb} showed that in
luminous sources such as \src, the most likely source for this BB-like
emission is the inner region of the accretion disk, from where the
reprocessed hard X-ray emission of the accretion column is
emitted. Such reprocessed emission also pulsates at the pulse period
of the hard X-ray component, most likely with a lower pulsed fraction
due to the large area where the reprocessing is taking place. This is
in agreement with both the change of the BB temperature and flux with
pulse phase (Figure~\ref{phaseResSpec}), and the slight decrease of
the PF at low energies compared to the high energies
(Figure~\ref{rmsVSene}). Under this assumption and for isotropic
emission, the inner radius of the disk is, $R_{\rm BB}^2=L_{\rm
  X}/(4\pi\sigma T^4)$ \citep{hickox04ApJ:bb}, where $\sigma$ is
the Stefan-Boltzmann constant, $L_{\rm X}$ is the non-thermal X-ray
luminosity, and $T$ is the BB temperature in K. The temperature we
calculate, however, is the apparent temperature of the plasma and is
related to the effective temperature through a color-correction (or
hardness) factor, $f_{\rm c}=T_{\rm c}/T_{\rm eff}$
\citep{damen90:colCorr,shimura95ApJ:colCorr,li05ApJS:KerrBB}, which is
usually taken to be between 1.5 and 2. Hence, the true inner radius of
the BB emission area is $R_{\rm in}=f_{\rm c}^2R_{\rm BB}$
\citep{kubota98PASJ:1009}. We find $R_{\rm in}=(4\pm1)\times10^{7}$~cm
(3$\sigma$ confidence), for $f_{\rm c}=1.8$. This is consistent with
the expected small accretion disk radius considering the low B field
of the source and its high luminosity.

The burst broadband spectrum requires only emission from the
non-thermal component, with fit parameters similar to the ones we
derive for the persistent emission. This reinforces our above picture
where we envisioned the burst emission to be the result of a sudden
increase of the mass accretion rate onto the neutron-star
pole. The non-detection of the BB component implies that the
reprocessing of the non-thermal emission may not have taken place
during bursts. This is possible, for instance, if the burst emission
is anisotropic away from the reprocessing material, i.e., the inner
accretion disk. Such anisotropy for the BP has already been discussed
by \citet{daumerie96Natur:1744} and \citet{nishiuchi99ApJ:1744} to
explain the extremely high luminosities of the bursts during the
previous two outbursts, which reached luminosities two orders of
magnitude above Eddington. This conclusion is also supported by the
timing properties of the source, for which the hard X-ray PF has been
seen to increase prominently during the bursts \citep{stark96ApJ:1744,
  woods00ApJ:1744}.

Finally, the broadband spectrum of the dip intervals is similar to
the persistent emission spectrum. The flux deficiency during dips is
primarily seen in the non-thermal component, where the {\sl cutoffPL}
flux decreased by $10\%$ compared to the persistent emission
flux. This is again in agreement with the accretion picture where the
dips are essentially the result of a fraction of the long-term
accreted matter not reaching the neutron star pole, instead replacing
the matter that produced the preceding burst.

\subsection{The Fe line complex}
\label{disFe}

The \asca\ observations during the first outburst of \src\ revealed a
feature between 6 and 8~keV in its persistent emission
spectrum. \citet{nishiuchi99ApJ:1744} modeled the spectrum with a
Gaussian line with a centroid energy of $6.7$ keV and an EW of
about 300~eV. The line was not resolved by the spectral resolution of
\asca, but its energy is indicative of a blend of emission lines from
different species. The \nustar\ persistent and dip spectra show a
similar emission excess at the same centroid energy and a somewhat
smaller EW (although consistent at the 3$\sigma$ level).

Using the \chandra\ HETGs we are able to resolve the broad feature
into three emission lines, which we identify as Fe~K from neutral
and/or lowly ionized species at $6.44\pm0.06$~keV, and highly 
ionized Fe~XXV and Fe~XXVI at $6.65_{-0.02}^{+0.01}$ and
$6.99\pm0.01$~keV (these are the best estimates of the line energies
from the dip$+$persistent emission spectrum, see also
\citealt{degenaar14:gro1744}).

We discuss first the Fe emission lines from highly ionized
species in the X-ray spectrum of \src. The gas producing the lines is
most likely photo-ionized by the X-ray emission of the neutron
star. In photo-ionized gas, He-like Fe emission lines are produced by
recombination and resonant scattering \citep{matt96MNRAS:Felines}, and
include four different transitions at slightly different energies, the
resonant line $w$, the two inter-combination lines $x$ and $y$, and
the forbidden line $z$ \citep[see e.g.,
][]{porquet00:Felines,porquet10:Felines}. Here, we could not resolve
the different resonances, however, from the centroid energy of the
Fe~XXV line, $6.65_{-0.02}^{+0.01}$~keV, we can safely conclude that the
resonant line $w$ (with mean energy at 6.700~keV) contributed
minimally to the line strength, hence the emission is
dominated by recombination \citep{bianchi02:Felines,
  bianchi05MNRAS:Felines,kallman01ApJS:xstar,matt96MNRAS:Felines}.

To investigate the origin of the highly ionized species, we simulate
XSTAR grids \citep{kallman01ApJS:xstar,bautista01ApJS:xstar} based on
the broadband X-ray spectrum of the source. We choose a covering
fraction of 0.2 (assuming an accretion disk), and solar abundances as
in \citet{grevesse96ASPC:abund}. Due to the high accretion rate of the
source, one would expect the photo-ionized gas to have a very large
density; hence, we examined different values of the gas density $n$
from $10^{10}$ to $10^{20}$~cm$^{-3}$, each time multiplying by 10, to
test the effects of density on reproducing the line shapes.  We find
that the best densities to reproduce the lines, resulting in
reasonable values of the ionization parameter ($1<\log\xi\, {\rm
  (erg~cm~s^{-1})}<5$), are 10$^{15}$ and 10$^{16}$~cm$^{-3}$. Here we
consider a density of $n=10^{16}$~cm$^{-3}$. Finally, we assumed no
turbulence in the gas. Fitting this XSTAR simulated grid to the
persistent+dip spectrum, we find a best fit value for the column
density in the gas\footnote{Due to the absence of absorption lines in
  the spectrum, the column density of the emitting gas could not be
  well constrained.} of $N_{\rm H}=3.4\times10^{22}$~cm$^{-2}$
($<4\times10^{23}$~cm$^{-2}$), and for the ionization parameter
$\log\xi\, {\rm (erg~cm~s^{-1})}=3.4_{-0.4}^{+0.8}$. The $\xi$
parameter is related to the total X-ray luminosity of the source, $L$,
the density of the ionized gas, $n$, and its distance from the
ionizing source, $R$, by $\xi=L/nR^2$. Solving for $R$, we find
$R=(2\pm1)\times10^{9}$~cm (3$\sigma$ confidence). This distance is
similar to the estimates of the ionized gas location in other sources
\citep[e.g.,
][]{ji09ApJ:herx1,paul05ApJ:gx14,jiminez05ApJ:HerX1,kallman03ApJ:0921},
and points towards reprocessing in an accretion disk corona.

The other interesting feature in our spectra is the (quasi$-$) neutral
Fe at 6.4~keV. There are three possibilities for the formation site of the
fluorescence Fe~K line in X-ray binaries: (i) a wind from the
companion seems to be unlikely in the case of \src, since the
companion is a low-mass star and accretion is most likely occurring
through Roche-lobe overflow \citep{finger96Natur:1744}; (ii) the
companion surface via reflection, which is also hard to achieve,
because of the very low inclination of the system \citep{
  finger96Natur:1744,rappaport97ApJ:1744} would result in a very low
EW for any Fe features \citep{basko78ApJ:FeK}; or (iii) the outer
regions of the accretion disk, by means of irradiation from the
central source. To test this third possibility, we fit a second XSTAR
grid, similar to the one above, to the dips+persistent spectrum. We
find that the (quasi$-$) neutral Fe line is well reproduced with an
ionization parameter $\log\xi\, {\rm (erg~cm~s^{-1})}=1.6$ ($<2.3$ at
3$\sigma$ confidence), much lower than the value required to model
the highly ionized lines. This ionization parameter corresponds to a
distance from the neutron star of $R=1.5\times10^{10}$~cm
($>7.0\times10^{9}$~cm at 3$\sigma$ confidence). This seems to
point to the outer regions of the accretion disk as the likely origin
of the Fe~K. Other X-ray binary sources showed, similar to \src,
Fe~K$\alpha$ lines most likely from the outer region of an irradiated
disk \citep[e.g.,][]{miller02ApJ:cygx1,reynolds10ApJ:0535}.

The 6.7~keV excess in the simultaneous \nustar\ and \chandra\ data is
also consistent with a broad line which we fit using a diskline model
(see also \citealt{degenaar14:gro1744}). The inner-disk radius that we
find ($R_{\rm in}=3_{-2}^{+5}\times10^{7}$~cm) is consistent with the
results of \citet{degenaar14:gro1744} and in agreement with a
magnetically truncated accretion disk. This radius, however, is more
than an order of magnitude smaller than the result we get from the
XSTAR fits to the highly ionized lines (assuming that the broad line
is consistent with Fe~XXV, \citealt{degenaar14:gro1744}). This could
either be due to the uncertainties in the density of the ionizing gas
and/or in the distance to the source, or the fact that other
broadening mechanisms, e.g., Compton scattering, are contributing to
the line profile.

% {\bf Both \nustar\ and \chandra\ data indicate a broad component in
%   the lines that is most likely associated with the Fe XXV feature
%   (see also \citealt{degenaar14:gro1744}). The inner-disk radius that
%   we find fitting the \nustar\ and \chandra\ data simultaneously
%   ($R_{\rm in}=3_{-2}^{+5}\times10^{7}$~cm) is consistent with the
%   results of \citet{degenaar14:gro1744}) and in-line with a
%   magnetically truncated accretion disk. This radius, however, is more
%   than an order of magnitude smaller than the result we get from the
%   XSTAR fits. This could either be due to the uncertainties in the
%   density of the ionizing gas and/or in the distance to the source, or
%   the fact that other broadening mechanisms, e.g., Compton scattering,
%   is contributing to the line profile.}

Due to the low statistics of the present observation, we could not
constrain any variations in the separate \chandra\ lines during dips
(Table~\ref{persisProp}). The \nustar\ Gaussian line fit to the
$6.7$~keV excess has energy, width, EW, and flux consistent within
$1\sigma$ between persistent and dip emission
(Table~\ref{persisProp}). The excess emission is not detected during
bursts, which means that either the super-Eddington burst X-ray
luminosity fully ionized the line-emitting region, including the Fe~K
region, or that the line strength remained more or less constant
during bursts, but was masked by the very bright continuum. To explore
the first possibility, we simulated the same XSTAR grid as above, but
instead of the persistent X-ray luminosity, we used the X-ray
luminosity as derived from the bursts. We fit this XSTAR table to
the burst spectrum and derived a 3$\sigma$ lower limit on
the ionization parameter $\log\xi\, {\rm (erg~cm~s^{-1})}>3.0$. This
limit represents the lowest ionization state that would result in the
featureless spectrum that we see during bursts. This lower limit
translates into an upper-limit on the radius of the ionized material
of $R<8.5\times10^{9}$~cm. Hence, the burst luminosity is capable of
fully ionizing the region of the disk where the highly ionized lines
are thought to originate ($(2\pm1)\times10^9$~cm). At the $3\sigma$
upper limit, it is capable of fully ionizing the region of the neutral
Fe ($>7.0\times10^9$~cm). This result does not exclude the second
possibility.

\subsection{Magnetic field estimate}

Similar to the previous two outbursts from \src, the increase in the
X-ray luminosity of the source is accompanied by an increase in the
spin period of the neutron star (GBM pulsar team, see footnote
2). This spin-up factor and persistent pulsed emission indicate that
the accretion onto the neutron star is not quenched at the
disk-magnetosphere boundary, i.e., the propeller effect is not
acting. For spin-up to occur during accretion, the inner disk
rotational frequency at the magnetospheric radius has to be greater
than the neutron star spin frequency, which results in an upper limit
on the magnetic dipole field of,

\begin{equation}
\footnotesize{
B<2~K^{-7/4}~(2\pi\nu)^{-7/6}~R^{-5/2}~L^{1/2}~(GM_{\rm NS})^{1/3},}
\end{equation}

\noindent where $L$ is the total X-ray luminosity assuming a distance
of 8~kpc, $G$ is the gravitational constant, $M$ and $R$ are the mass
and radius of the neutron star, taken to be 1.4~${\rm M_{\odot}}$ and
10~km, $\nu$ is the neutron star spin frequency, and $K$ is a
dimensionless parameter between 0.5 and 1 \citep{ghosh79ApJ:Kparam,
  spruit93ApJ:accDisk,arons93ApJ:magFields,ostriker95ApJ,
  wang96ApJ:kparam,finger96ApJ:A0535}. We find
$B<3.5\times10^{11}\, K^{-7/4}$~G. This value is consistent with
the estimates of the previous two outbursts \citep[e.g.,][]{
  finger96Natur:1744,bildsten97ApJ:1744}. We note that the true
upper-limit is lower than the above derived value since the source
started spinning-up at earlier stages in the outburst when the source
luminosity was lower.

Assuming that the BB component is the result of reprocessing in the
inner regions of the accretion disk, we could also use the BB radius
estimate (Section~\ref{discussBBs}) to derive the strength of the
dipole field of the source. The inner accretion disk radius can be
written as

\begin{equation}
\footnotesize{
r_{0}=K~\mu^{4/7}~(GM)^{1/7}~R^{-2/7}~L^{-2/7},}
\end{equation}

\noindent with $B=2\mu R^{-3}$. We find

\begin{equation}
\footnotesize{
B=9^{+1}_{-2}\times10^{10}~\left(\frac{K}{1}\right)^{-7/4} \left(\frac{R}{10~{\rm km}}\right)^{-5/2} \left(\frac{M}{1.4~{\rm M_\odot}}\right)^{-1/4} \left(\frac{D}{8~{\rm kpc}}\right)~{\rm G}}
\end{equation}

\noindent  which is consistent with the above upper-limit and the
expected low dipole field of the source.

\section{Conclusion}

We studied the broad-band X-ray emission ($0.5-70$~keV) of the BP
from a $\simeq3$~h simultaneous \nustar-\chandra\ observation during
its third detected outburst since discovery and after nearly 18 years
of quiescence. These data were taken a few days before the outburst
reached its peak.

A total of seven bursts are detected during our observation. Temporal 
analysis revealed that the first six bursts have comparable shapes,
consisting of a single pulse with duration of $12$~s, and a faster
rise than decay time. The last burst has a double-peaked morphology
with a duration of about 25~s. All seven bursts, however, have equal
fluences with an average of about $10^{-6}$~erg~cm$^{-2}$. Similar to
previous results, we find an average ratio of the burst to the
persistent emission fluence $\alpha\approx10$ (with the exception of
one burst where $\alpha=30$), pointing to the type~II origin for the
bursts. Each of the seven bursts is followed by a dip in the
persistent emission flux, which recovers exponentially with a
characteristic time-scale $\tau\approx190$~s. We find an average
missing fluence in the dip of about $10^{-6}$~erg~cm$^{-2}$,
consistent with the fluence emitted in the bursts. This indicates that
the energy emitted during the burst is compensated for in the dip, and
that the long-term accretion rate is constant. The pulse-profiles of
the persistent and the dip intervals are nearly sinusoidal with only
weak contribution from the second harmonic. The PF increases from about
10\% at 4~keV to 15\% at 13~keV, and remains constant thereafter.

The BP persistent and dip broadband spectra are identical and well
fit with a BB with $kT=0.5$~keV, a {\sl cutoffPL} with an index
$\Gamma=0.0$ and an energy rolloff $E_{\rm fold}=7$~keV, a 10~keV
feature assumed to be the result of inadequate modeling of the
{\sl cutoffPL}, and a 6.7~keV emission feature, all modified
by neutral absorption. Phase-resolved spectroscopy shows that the BB
and the {\sl cutoffPL} components show variations at the pulse period
of the source, both getting harder at pulse maximum, whereas no
significant changes are seen in the 10~keV and the 6.7~keV feature.

Assuming that the BB is reprocessing of the non-thermal emission in
the inner regions of the accretion disk, we derive an inner disk
radius $R=4\times10^7$~cm. This radius translates into a dipole
magnetic field of $B\approx9\times10^{10}$~G.

The \chandra/HETG spectrum resolved the 6.7~keV feature into
(quasi-)neutral and highly ionized Fe~XXV and Fe~XXVI narrow emission
lines. Modeling the highly ionized lines with XSTAR places the
emitting region at a distance of about $10^9$~cm from the neutron
star, consistent with an accretion disk corona origin. Using a similar
XSTAR  grid to model the (quasi-) neutral Fe, we find that it
originates from a distance $\gtrsim10^{10}$~cm, most likely the outer
regions of an accretion disk.

The broadband burst spectrum, with a peak flux more than an order of
magnitude higher than Eddington, is well fit with a {\sl cutoffPL}
and a 10~keV feature, with similar fit values compared to the
persistent and dip spectra. The burst spectrum, however, lacks a
thermal component (BB) and Fe features. If the burst emission were
anisotropic (beamed), the lack of the BB component is expected since
no reflection of the burst photons on the inner disk would take
place. Similarly the Fe XXV, FeXXVI, and the neutral Fe lines would
remain at the flux levels detected in the persistent and dip emission
and, therefore, are too weak to be detected above the strong burst
continuum. If, on the other hand, the burst emission is isotropic, we
show that the disk region where the Fe XXV and FeXXVI lines would be
produced is now fully ionized; the neutral iron line could still be
at very low levels and masked by the continuum. In that case, however,
we would expect a strong BB component, which is not detected. We
conclude that, as suggested by \citet{daumerie96Natur:1744} and
\citet{nishiuchi99ApJ:1744}, the burst emission is highly beamed.

\section*{Acknowledgments}
This work was supported under NASA Contract No. NNG08FD60C, and made use of data
from the {\it NuSTAR} mission, a project led by  the California Institute of
Technology, managed by the Jet Propulsion  Laboratory, and funded by the National
Aeronautics and Space Administration. We thank the {\it NuSTAR} Operations,
Software and  Calibration teams for support with the execution and analysis of
these observations.  This research has made use of the {\it NuSTAR}  Data
Analysis Software (NuSTARDAS) jointly developed by the ASI  Science Data
Center (ASDC, Italy) and the California Institute of  Technology (USA).

\end{document}